%% file: RadComPow_double_column.tex
\documentclass[10pt]{IEEEtran}
\usepackage{amsmath}
\usepackage{amssymb}
\usepackage{amsfonts}  
\usepackage[short]{optidef}
\allowdisplaybreaks[4]
\usepackage[subtle]{savetrees}

\usepackage{amsthm}
\usepackage{enumitem}

\usepackage{graphicx}
\usepackage{float}  
\usepackage{tikz,lipsum}
\usepackage[labelformat=simple]{subcaption}
\DeclareCaptionLabelSeparator{periodspace}{.\quad}
\usepackage{comment}
\usepackage{caption}
\usepackage{pdfpages}   
\usepackage{epstopdf}
\usepackage{bbm}
\usepackage{cite}
\usepackage{url} 

\usepackage[ruled,vlined]{algorithm2e}  
\usepackage[normalem]{ulem}
\newcommand{\areal}[1]{{#1}^{\mathrm{R}}}
\newcommand{\aimag}[1]{{#1}^{\mathrm{I}}}

\usepackage{mathtools}
\usepackage{setspace}
\include{notation_new} 

\begin{document}

\title{\date{} Multi-functional OFDM Signal Design for Integrated Sensing, Communications, and Power Transfer}
\author{Yumeng Zhang, Sundar Aditya,~\IEEEmembership{Member,~IEEE} and Bruno Clerckx,~\IEEEmembership{Fellow,~IEEE} 
\thanks{The authors are with the Department of Electrical and Electronic Engineering, Imperial College London, London SW7 2AZ, U.K. (e-mail:
\{yumeng.zhang19, s.aditya, b.clerckx\}@imperial.ac.uk)}
\thanks{B. Clerckx is also with Silicon Austria Labs (SAL), Graz A-8010, Austria.}
}
\maketitle

\begin{abstract}
The wireless domain is witnessing a flourishing of integrated systems, e.g. (a) integrated sensing and communications, and (b) simultaneous wireless information and power transfer, due to their potential to use resources (spectrum, power) judiciously. Inspired by this trend, we investigate integrated sensing, communications and powering (ISCAP), through the design of a wideband OFDM signal to power a sensor while simultaneously performing target-sensing and communication. To characterize the ISCAP performance region, we assume symbols with non-zero mean asymmetric Gaussian distribution (i.e., \textit{the input distribution}), and optimize its mean and variance at each subcarrier to maximize the harvested power, subject to constraints on the achievable rate (communications) and the average side-to-peak-lobe difference (sensing). The resulting input distribution, through simulations, achieves a larger performance region than that of (i) a  symmetric complex Gaussian input distribution { with identical mean and variance for the real and imaginary parts}, (ii) a zero-mean symmetric complex Gaussian input distribution, and (iii) the superposed power-splitting communication and sensing signal (the coexisting solution). In particular, the optimized input distribution balances the three functions by exhibiting the following features: {(a) symbols in subcarriers with strong communication channels have high variance to satisfy the rate constraint, while the other symbols are dominated by the mean, forming a relatively uniform sum of mean and variance across subcarriers for sensing; (b) with looser communication and sensing constraints,  large absolute means appear on subcarriers with stronger powering channels for higher harvested power.} As a final note, the results highlight the great potential of the co-designed ISCAP system for further efficiency enhancement.
\end{abstract}

\begin{IEEEkeywords}
Multi-functional OFDM waveform design, Integrated Sensing, Communications and Powering (ISCAP), Integrated Sensing and Communication (ISAC), Wireless Information and Power Transfer (WIPT)
\end{IEEEkeywords}

\section{Introduction}
\subsection{Background} 
Future wireless networks are expected to feature billions of devices carrying out tasks like sensing, communications and computing. This calls for better spectrum utilization to accommodate more users and services, as well as a sustainable energizing technique \cite{paul2016survey,2013Joint,zhang2013mimo}. For the former, integrated sensing and communications (ISAC) using a common transmit signal has been widely investigated \cite{wymeersch2021integration,yuan2021integrated,10004202}. For the latter, thanks to the consistent decrease in the power consumption of computational tasks \cite{koomey2010implications}, wireless information and power transfer (WIPT) has been regarded as a promising technique to provide sustainable powering for low-power sensors, by generating power from the communicating RF signals.

The literature on ISAC and WIPT has shown that a larger performance region (i.e., communication-sensing (C-S) region for ISAC, and a communication-powering (C-P) region for WIPT) can be realized with properly co-designed signals, as opposed to superposed signals in a coexisting scenario that combines the communications and sensing/powering signals through time-splitting/power-splitting. Inspired by this, a question then arises: can we achieve (a suitably-defined) better performance by realizing sensing, communications and powering using a common signal, namely integrated sensing, communications and powering (ISCAP)? \textcolor{blue}{If achievable, ISCAP can play a crucial role in future networks by reducing the power burden and enhancing spectrum efficiency, especially in environments with densely deployed low-power sensors and massive connectivity demands. A possible use case is ubiquitous sensing requiring a large number of sensors to communicate, sense and be autonomously powered, for applications in as smart homes, smart city, transportation, automotive, agriculture, logistics, emergency, security, prevention, and defense.} To answer the proposed question, we need to: a) decide and design a signal that is capable of performing sensing, communications and powering simultaneously and satisfyingly, b) evaluate the performance gain of the ISCAP signal using the proposed input signal over the traditional co-existing scenarios, c) identify the rationale of the performance gain by analyzing how the optimal signal trades off between the three functions. 

 {This paper first explores the reason for using OFDM signals in ISCAP and the need for optimizing the input distribution (i.e., the probability distribution of the communication symbols (CS)) over the subcarriers in the literature review. Then, our contributions in this paper address points a), b) and c) in detail.}

\subsection{Literature review} 
\textcolor{blue}{The literature on wireless networks has collectively highlighted OFDM's significant potential as a promising wideband signal for the multi-functional ISCAP system. The reasons are threefold. Firstly, OFDM has been widely employed in practice (e.g., IEEE 802.11 for wireless local area networks \cite{crow1997ieee}, IEEE 802.16 for wireless metropolitan area networks \cite{hoymann2005analysis}, and also in 4G and 5G \cite{7469313}) and intensively explored for sophisticated transceiver design
\cite{speth2001optimum}. Consequently, OFDM is more implementation-friendly if adopted as the initial waveform in ISCAP exploration. Secondly, OFDM shows satisfying performance for all ISCAP functionalities. Specifically, OFDM exhibits good auto-correlation properties that are advantageous in sensing, together with an efficient 2-dimensional Fast-Fourier-Transform (2D-FFT) radar processor \cite{sturm2011waveform}. Moreover, OFDM provides robustness against multipath fading for communications \cite{hwang2008ofdm}, and offers high peak-to-average-power-ratio properties that are beneficial in wireless power transfer (WPT) \cite{clerckx2016waveform}. Thirdly, the waveform design for OFDM  has been extensively studied in both ISAC and WIPT \cite{aubry2016optimization,5599316, donnet2006combining,barneto2019full,xu2020multi,varshney2008transporting,clerckx2018fundamentals}, which facilitates our extension of the optimal OFDM waveform design to the ISCAP system. In the following, we will summarize the intuitive signal design in ISAC and WIPT literature.}

Ample research has been conducted on optimizing the OFDM waveform in ISAC, specifically the power allocation across OFDM subcarriers to achieve a better C-S  region that depicts the optimal achievable sensing performance given a communication constraint \cite{aubry2016optimization,5599316, donnet2006combining,barneto2019full,xu2020multi}. Towards that, various metrics have been developed to evaluate the sensing performance when using OFDM signals, such as the detection/false alarm probability (FAP) \cite{sen2010adaptive,hsu2021analysis}, the ambiguity function \cite{sturm2011waveform}, the spectrum matching error \cite{cheng2021hybrid} 
, the radar mutual information \cite{sen2010ofdm,ouyang2023mimo} and the Cramer-Rao Bound (CRB) \cite{bicua2018radar}. 

Regarding WIPT, the performance is characterized by the C-P region, which represents the maximal harvested power given the communication constraint \cite{varshney2008transporting,clerckx2018fundamentals,6623062}
. Successive efforts have been made to enlarge the C-P region in WIPT \cite{son2014joint,park2014joint,nasir2013relaying}, through which the significance of modelling the non-linearity of the energy harvester/the rectenna at the harvesting receiver was uncovered and shown to exert a fundamental effect on the optimal signal design \cite{xiong2017rate,boshkovska2017robust,clerckx2016waveform,huang2017large,abeywickrama2021refined,clerckx2018toward,kim2020signal,clerckx2017wireless,varasteh2019swipt}. \textcolor{blue}{In particular, accounting for the energy harvester's non-linearity, \cite{clerckx2017wireless} examined the multi-carrier signal in WIPT and showed that the modulated signal  {(carrying  information with CSCG input)} superposed by the unmodulated signal (deterministic without information) can enlarge the C-P region, with the former necessary for communications and the latter beneficial for powering. }

 {\cite{clerckx2017wireless} motivates the consideration of the impact of random CSs on powering in OFDM WIPT. As a follow up to \cite{clerckx2017wireless}, \cite{varasteh2019swipt} optimizes the input distribution of the OFDM CSs, i.e., the  {symbol} mean (\textcolor{blue}{as a counterpart of the unmodulated signals of the superposition multi-carrier waveform in \cite{clerckx2017wireless}}) and the symbol variance (\textcolor{blue}{as a counterpart of the modulated signals of the superposition multi-carrier waveform in \cite{clerckx2017wireless}}) of each sub-carrier assuming Gaussian distribution, and achieves a larger WIPT C-P region than \cite{clerckx2017wireless} by assuming asymmetric Gaussian input. The findings in WIPT further intrigue similar considerations for enlarging the C-S region in ISAC,  where the OFDM symbols are also inherently random to carry communication information and will affect sensing's performance\cite{2022ISAC}. } 
The optimal input distribution in \cite{2022ISAC}/\hspace{1sp}\cite{varasteh2019swipt} not only enhances the performance region in ISAC/WIPT but also uncovers fundamental trade-offs between different functions, namely, (a) for ISAC \cite{2022ISAC}, the randomness of CS magnitudes degrades sensing performance while enhancing communication rates. Hence, the optimal input distribution trades off between (i) sensing's preference for allocating power uniformly to the symbol \emph{mean} of each subcarrier and (ii) communication's preference for  { allocating power to the symbol \emph{variance} across subcarriers in a water-filling way}; (b) for WIPT \cite{varasteh2019swipt},  {powering also prefers high absolute symbol means similarly to sensing, and a further performance region gain is observed when using asymmetric non-zero mean complex Gaussian distribution (i.e., featuring different means/variance between the real and imaginary parts of the complex Gaussian). 

\textcolor{blue}{Recently, \cite{chen2022isac} established the first narrowband MIMO ISCAP system where the authors optimized the transmit beam pattern and achieved a better overall performance than the power/time-splitting signals in the co-existing scenario \cite{10118850}.  Considering the necessity of wideband signals for higher range resolution in sensing, this paper investigates wideband OFDM signals in ISCAP that performs range-velocity estimation for sensing, compared with \cite{chen2022isac} where angle estimation is the focus. Moreover, we consider a more realistic non-linear energy harvester model for powering when designing the OFDM signals. Besides validating the effectiveness of the designed signal in ISCAP, this paper also studies the spectrum interaction between the three functionalities and reveals the general trend of the optimal input distribution of the OFDM signal in the spectrum.}

\subsection{Contributions}
Inspired by the conclusions in WIPT and ISAC \cite{2022ISAC,varasteh2019swipt}, we design the OFDM CSs' input distribution across subcarriers in the spectrum domain, with an assumption of a non-zero mean asymmetric Gaussian distribution. Our contributions are summarized as follows:
\begin{enumerate}
\item  We set up a wideband ISCAP system, where an OFDM signal is transmitted to power a sensor, communicate with an information decoder and sense a point target simultaneously. To evaluate the system performance region, we develop the metric of each function $\textendash$ respectively the average integrated-side-to-peak-lobe-difference (aISPLD) for sensing\footnote{The aISPLD is proposed as a scaling term of the upper bound of the average FAP in OFDM ISAC in \cite{2022ISAC}.}, the achievable rate for communications, and the harvested power for powering $\textendash$ accounting for CS  randomness. For the power metric in particular, we follow the non-linear energy harvester model in \cite{clerckx2016waveform} and derive the average harvested power of cyclic prefix (CP) OFDM (CP-OFDM).  {This is the first paper to model a wideband ISCAP system and the first paper to consider the power contribution from the OFDM CP part where inter-symbol interaction is involved when designing the OFDM symbols\footnote{ {Different from \cite{kassab2022superposition} which reconstructs the OFDM CP by superposing rectangular pulses with the traditional CP (circular-shift of the data) for better power harvesting, this paper accounts for the traditional CP construction only and evaluates the impact of the CS input distribution on the harvested power from the CP part, which is emitted in \cite{varasteh2019swipt} where only the data part is considered. Hence, the paper will not encounter degraded communication performance caused by the additional rectangular pulses in CP as in \cite{kassab2022superposition}.}}. }

\item  Based on the metrics of the three functions, we construct an optimization problem w.r.t the symbol mean and symbol variance of each OFDM subcarrier to maximize the harvested power with constraints on achievable rate and aISPLD.  The optimization problem is solved using the alternating direction method of multipliers (ADMM). During each ADMM iteration, we first solve the total allocated power at each subcarrier (the sum of the mean and variance)  by using successive convex approximation (SCA), after which we determine the allocation of power over the mean and variance at each subcarrier by utilizing the parametric successive convex approximation (PSCA).

\item  Through simulations, we verify the performance region gain of the proposed input distribution in ISCAP over the co-designed symmetric/zero-mean input distribution as well as the power-splitting input signals in a coexisting scenario. We also gain insight into the fundamental trade-offs between the three functions on input distribution, which can be summarized as follows: a) powering prefers allocating power { to the symbol mean with asymmetric real and imaginary part in proportional to the corresponding's powering channel}, sensing prefers allocating power {to the symbol mean uniformly across subcarriers}, and communication requires allocating power to the  {symbol} variance in proportion to communication channels' strength of its subcarrier; b) In a general ISCAP set-up with communication and sensing constraints,  a part of subcarriers has a high variance for achievable rate (and usually a low mean) while the remaining subcarriers are dominated by the mean, forming a relatively uniform power allocation across subcarriers for the aISPLD constraint. Given looser constraints on achievable rate and aISPLD, the optimal input distribution shifts to a power-favouring distribution with asymmetric mean allocation adaptive to the powering channels.
\end{enumerate}

\subsection{Organization}  Section \ref{sec_SSModel} models a single antenna OFDM ISCAP system with a point target, where the signal at each stage is mathematically expressed. The metrics for powering, sensing and communications are also expressed based on an asymmetric non-zero mean Gaussian input. Then, Section \ref{section_opt} optimizes the mean and the variance of the input distribution across subcarriers, by maximizing the harvested power constraining on the achievable rate and the aISPLD. Section \ref{sec_sim} provides simulation results and Section \ref{sec_con} draws the conclusion. 

\subsection{Notation}  
Throughout the paper, matrices and vectors are respectively denoted by bold upper case and bold lower letters. $\mathfrak{R/I}\{x\}$ denotes the real/imaginary part of the complex number $x$.  $|{x}|$ represents the amplitude of the complex scaler ${x}$ and $\|\mathbf{x}\|$ represents the $l_2$ norm of vector $\mathbf{x}$. For a vector $\mathbf{x}$ (matrix $\mathbf{X}$), ${x}_k$  ($X_{k,m}$) is its $k^{\mathrm{th}}$  ($k^{\mathrm{th}}$ row, $m^{\mathrm{th}}$ column) entry. $\mathbf{I}_{K}$ represents a $K\times K$ identity matrix, and $\mathbf{1}_K$ represents an all-one vector with dimension $K\times~1$ - the subscript is omitted when the dimension is clear. $(\cdot)^H$,  $(\cdot)^T$ and $\text{Tr}(\mathbf{X})$ denote the Hermitian, transpose and trace operators respectively. $\otimes$ is the Kronecker product. $\delta{(x)}$ is the delta function. $\text{diag}\left( \mathbf{A}\right)$ denotes the vector of diagonal entries of $\mathbf{A}$.  Similarly, $\text{diag}\left( \mathbf{a}\right)$ denotes the diagonal matrix formed by $\mathbf{a}$. $\left<n\right>_k$  denotes $n$ modulo $k$.  The discrete Fourier transform (DFT) for sequences $\{x[n]\}$/vector $\{\mathbf{x}\}$ is denoted by $\text{DFT}\{x[n]\}$/$\text{DFT}\{\mathbf{x}\}$ with its $k^{\mathrm{th}}$ entry being $\text{DFT}\{x[n]\}_{(k)}$/$\text{DFT}\{\mathbf{x}\}_{(k)}$, and similarly for the inverse DFT (IDFT). \textcolor{blue}{$\mathbf{f}_{n}=\left[1,~e^{j\frac{2\pi n}{K}},~...,~e^{j\frac{2\pi n(K-1)}{K}}\right]$ denotes the $n^{\mathrm{th}}$ row of the IDFT matrix.}  {$f_{\mathrm{I}}(x\in \kappa)$ is the indicator function of $x$ belonging to set $\kappa$.}  $\mathcal{E}\{x(t)\}$ denotes the time average of a signal $x(t)$ and $\mathbb{E}_X\{f_X(x)\}$ is the expected value of $f_X(x)$ over the distribution of the random variable $X$ ($X$ is omitted after the first clarification). 

\begin{ndef}[Real Gaussian Distribution]
$\ncalN(\mu, \sigma^2)$ denotes the real Gaussian distribution with mean $\mu$ and variance $\sigma^2$.
\end{ndef}

\begin{ndef}[Complex Gaussian Distribution]
\label{def:complex_gaussian}
Let $X_R \sim \ncalN(\mu_R, \sigma_R^2)$ and $X_I \sim \ncalN(\mu_I, \sigma_I^2)$ denote a pair of \underline{independent} real Gaussian random variables. Then, $X = X_R + jX_I$ is said to have a complex Gaussian distribution. 
\end{ndef}

\begin{ndef}[Symmetric and Asymmetric Complex Gaussian Distribution]\label{def:complex_gaussian_symmetric}
In Definition~\ref{def:complex_gaussian}, if $\mu_R = \mu_I = \mu$ and $\sigma_R^2 = \sigma_I^2 = \sigma^2/2$, then $X$ is said to have a \underline{symmetric} complex Gaussian distribution\footnote{\label{footnote_symmetric} {Typically, a symmetric complex Gaussian distribution only requires the variance of the real and imaginary parts to be identical \cite{clerckx2017wireless,varasteh2019swipt}. In Definition \ref{def:complex_gaussian_symmetric}, we also require the means of the real and imaginary parts to be equal. This is because the (complex) mean has a significant bearing on the harvested power and sensing performance, and we wish to capture the performance difference between having identical/different values for the real and imaginary parts of the mean, which will be stated in the simulations in detail.}}, denoted by $\ncalC \ncalN(\mu, \sigma^2)$. The special case of $\ncalC \ncalN(0, \sigma^2)$ is referred to as the circularly symmetric complex Gaussian (CSCG) distribution. In all other cases, $X$ is said to have an \underline{asymmetric} complex Gaussian distribution.
\end{ndef}

\section{System and Signal Models}
\label{sec_SSModel}
This section begins by modelling the ISCAP system and the transmit
OFDM signal, as shown in  Fig. \ref{fig_system_model}. We then identify performance metrics for each function, namely, the harvested power per OFDM symbol at the energy harvester  in Section \ref{section_powering}, the average FAP (approximated by the aISPLD) for range-velocity estimation at the sensing transceiver in Section \ref{section_radar}, and the achievable rate per OFDM symbol at the communication receiver in Section \ref{section_communication}.
\begin{figure*}[ht]
\includegraphics[scale=0.45]{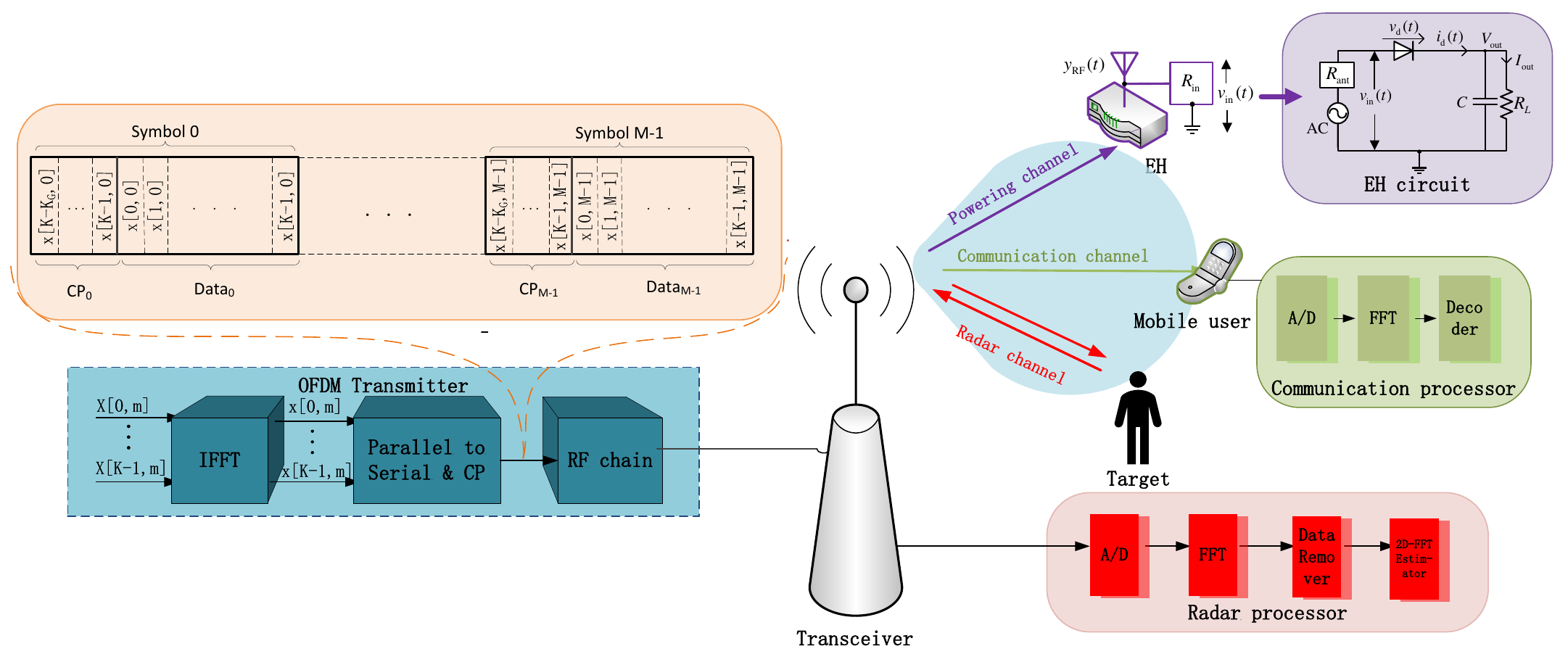}
\caption{\textcolor{blue}{System and OFDM Signal Model for ISCAP.}}
\label{fig_system_model}
\end{figure*}

\subsection{Transmit Signal}
\label{section_transmit_signal}
Without loss of generality, we consider a single antenna ISCAP system as shown in Fig. \ref{fig_system_model}, with a multi-functional transmitter and separated receivers for sensing, communications and powering, among which the sensing receiver is co-located with the transmitter. Besides the block model of the  ISCAP system, Fig. \ref{fig_system_model} also depicts the OFDM symbol model at the transmitter, where $M$ OFDM symbols with $K$ orthogonal subcarriers and $K_G$ CP sub-pulses for each OFDM symbol are used.

The CS at the $k^{\mathrm{th}}$ ($k=0,~\cdots,~K-1$) subcarrier of the $m^{\mathrm{th}}$ ($m=0,~\cdots,~M-1$) OFDM symbol can be expressed as
\begin{align}
\label{eq_transmit_scaler}
X[k,~m]\overset{\vartriangle}{=}\areal{X}[k,~m]+j\aimag{X}[k,~m],
\end{align}
where $\areal{X}[k,~m]\sim\mathcal{N}(\areal{\mu}_{k},~{\areal{\sigma}_{k}}^2)$ and $\aimag{X}[k,~m]\sim\mathcal{N}(\aimag{\mu}_{k},~{\aimag{\sigma}_{k}}^2)$  (assume the same input distribution spectrum across OFDM symbols). Hence, the average power on the real and imaginary part of the $k^{\mathrm{th}}$ subcarrier is given as
\begin{subequations}\label{eq_Gaussian_power_define}\begin{align}
\areal{p}_{k}\overset{\vartriangle}{=}&\mathbb{E}_{X^\mathrm{R}[k,~m]}\left\lbrace {X^{\mathrm{R}}[k,~m]}^2 \right\rbrace={\areal{\mu}_{k}}^2+{\areal{\sigma}_{k}}^2,~\forall ~m,\\
\aimag{p}_{k}\overset{\vartriangle}{=}&\mathbb{E}_{X^\mathrm{I}[k,~m]}\left\lbrace {X^{\mathrm{I}}[k,~m]}^2 \right\rbrace={\aimag{\mu}_{k}}^2+{\aimag{\sigma}_{k}}^2,~\forall ~m.
\end{align}\end{subequations}

After IDFT, we obtain the time domain signal $x[n,~m]$ ($n=0,~\cdots,~ K-1$), namely the $n^{\mathrm{th}}$ sub-pulse of the $m^{\mathrm{th}}$ symbol, which, after adding the CP, becomes  
\begin{equation}
\label{eq_CP_discrete2}
x_{\mathrm{CP}}[n,~m]= x[\left<n-K_G\right>_K,~m], ~n=0,~\cdots,~ K'-1,
\end{equation} 
where $K'=K_G+K$.

Then, the baseband signal is 
\begin{subequations}
 \begin{align}
\label{eq_transmit_signal}
x(t)=&\sum_{m=0}^{M-1}\sum_{n=0}^{K'-1} x_{\mathrm{CP}}[n,~m]\text{sinc}\left[Bt-(n+mK')\right]\\
\overset{\bigtriangleup}{=} &\sum_{q=0}^{K'M-1} x_{\mathrm{CP}}[q]\text{sinc}\left[Bt-q\right],~(q=mK'+n),
\end{align}   
\end{subequations}
with $x_{\mathrm{CP}}[q]=x_{\mathrm{CP}}[n,~m]$ and $B$ is the bandwidth of the OFDM signal.

After up-converting, the transmit signal is expressed as 
\begin{align}
\label{eq_transmit_signal_rf}
x_{\mathrm{RF}}(t)&=\sum_{m=0}^{M-1}\sum_{n=0}^{K'-1} x_{\mathrm{CP}}[n,~m]\text{sinc}\left[Bt-(n+mK')\right]e^{j2\pi f_c t},
\end{align}
where $f_c$ is the central frequency.

\textcolor{blue}{For future analysis, we define the following vectors}
\begin{subequations}\color{blue}
\begin{align}
\label{eq_vector_X}
\mbox{Symbol vector:}~\mathbf{x}_m\overset{\vartriangle}{=}&\left[ X[0,~m] ,~\cdots,~ X[K-1,~m] \right]^T,\\
        \mbox{Symbol real vector:}~\mathbf{\overline{x}}_m\overset{\vartriangle}{=}&\left[\mathfrak{R}\{ \mathbf{x}_m\}^T, ~\mathfrak{I}\{\mathbf{x}_m\}^T\right]^T,\\
\label{eq_x_avector}
\mbox{Symbol mean vector:}~\boldsymbol{\mu}\overset{\vartriangle}{=}&\left[\areal{\mu}_{0},~\areal{\mu}_{1},~\cdots,~\areal{\mu}_{K-1},~\aimag{\mu}_{0},~\cdots,~\aimag{\mu}_{K-1}\right]^T,\\
\label{eq_x_sigma}
\mbox{Symbol variance vector:}~\boldsymbol{\sigma}\overset{\vartriangle}{=}&\left[{\areal{\sigma}_{0}}^2,~{\areal{\sigma}_{1}}^2,~\cdots,~{\aimag{\sigma}_{0}}^2,~\cdots,~{\aimag{\sigma}_{K-1}}^2\right]^T,\\
\label{eq_x_p}
\text{Symbol power vector:}~\mathbf{p}\overset{\vartriangle}{=}&\left[\areal{p}_{0},~\areal{p}_{1},~\cdots,~\areal{p}_{K-1},~\aimag{p}_{0},~\cdots,~\aimag{p}_{K-1}\right]^T\\
\label{eq_define_U}
\text{Symbol mean matrix:} ~ \mathbf{U}\overset{\vartriangle}{=}&\boldsymbol{\mu}\boldsymbol{\mu}^T \\
\label{eq_define_P}
\text{Power allocation matrix:} ~ \mathbf{P}\overset{\vartriangle}{=}&\mathbf{U}+\text{diag}\left(\boldsymbol{\sigma} \right),
\end{align}
\end{subequations}
where $\mathbf{U}$ and $\mathbf{P}$ are defined for Section \ref{sec_sim}. Therein, we will optimize the ISCAP system performance w.r.t $\mathbf{U}$ and $\mathbf{P}$ to decide $\boldsymbol{\mu}$ and $\boldsymbol{\sigma}$\footnote{\textcolor{blue}{The input distribution is optimized w.r.t both the real and imaginary parts of the OFDM CSs, simultaneously reflecting the behaviour of their amplitude and phase.}}.

\subsection{Metric for Powering}
\label{sec_Metric}
\label{section_powering}
This section models the received signal at the energy harvester, and derives the corresponding scaling term of the harvested power as the powering metric. 

Assume that the signal is transmitted over a multipath channel with $L$ paths, where the gain and delay of the $l^{\mathrm{th}}$ $(0\leq ~l\leq~L-1)$ tap are respectively $a_l$ and $\tau_l$. \textcolor{blue}{We assume a static powering channel, whose channel state information (CSI) has been perfectly estimated in advance through pilot signals}. The real received signal at the energy harvester is
\begin{align}\label{eq_real_y(t)}
y_{\mathrm{RF}}(t)=&\sqrt{2}\mathfrak{R}\left\lbrace \sum_{l=0}^{L-1} a_l x_{\mathrm{RF}}(t-\tau_l)+w(t)\right\rbrace,
\end{align}
where $w(t)\sim\mathcal{CN}(0,~\sigma_{\mathrm{n_P}}^2)$ is the additive white Gaussian noise (AWGN) at the energy harvester.

At the energy harvester, the impinging signal $y_{\mathrm{RF}}(t)$ is converted into direct current (DC) via a rectenna circuit for power harvesting. We model the non-linear rectenna referring to \cite{clerckx2016waveform}, \textcolor{blue}{whose equivalent circuit is depicted in Fig. \ref{fig_system_model} on the top right. Specifically, in  Fig. \ref{fig_system_model}, $y_{\mathrm{RF}}(t)$ is picked up at the receiving antenna, which serves as a voltage source with average power $\mathcal{E}\left\lbrace {y}_{\mathrm{RF}}(t)^2\right\rbrace  $ and an inner impedance $R_\mathrm{ant}$. The rectenna circuit in Fig. \ref{fig_system_model} is composed of a diode and a low pass filter (LPF)\footnote{\textcolor{blue}{The LPF is modelled as the simplest capacitor-resistor circuit in Fig. \ref{fig_system_model}. However, for future analysis, we assume that
the LPF filters out all the high-frequency harmonic components in $i_\mathrm{d}(t)$ ideally.}} with $R_\mathrm{in}$. We assume $R_\mathrm{in}=R_\mathrm{ant}$.   Then, the input voltage fed into the rectenna is $v_{\mathrm{in}}(t)={y}_{\mathrm{RF}}(t)\sqrt{R_\mathrm{ant}}$. On this basis, the current at the diode, assuming the small signal model \cite{clerckx2016waveform,clerckx2017wirelessly,varasteh2019swipt}, is $i_d(t)=i_s\exp\left\lbrace    v_d(t)/(nv_t)-1\right\rbrace $ where $v_d(t)=v_{\mathrm{in}}(t)-V_{\mathrm{out}}$  with $ V_{\mathrm{out}}$ being the output DC voltage of the rectenna. We simplify the expression of $i_d(t)$ by  operating Taylor approximation at point $-V_{\mathrm{out}}$, which gives $i_d(t)=\sum_{i=0}^{\infty}k_i v^i_{\mathrm{in}}(t)=\sum_{i=0}^{\infty}k'_i R^{\frac{i}{2}}_{\mathrm{ant}} {y}_{\mathrm{RF}}(t)^i$ with $k'_i$ being the $i^{\mathrm{th}}$ order Taylor coefficient of the diode transfer characteristics.  Then, $i_d(t)$ passes through the LPF where its high-frequency components are filtered out before being fed into the load $R_{\mathrm{L}}$. Consequently, the current at the load $R_{\mathrm{L}}$ is $I_{\mathrm{out}}=\mathcal{E}\left\lbrace i_d(t)\right\rbrace=\sum_{i=0\text{, even}}^{\infty}k_i\mathcal{E}\left\lbrace{y}(t)^i\right\rbrace$. More details are in \cite{clerckx2016waveform}. To maximize the generated power at the load $R_{\mathrm{L}}$ is equivalent to maximizing the current at the load $R_{\mathrm{L}}$, i.e., $I_{\mathrm{out}}$. In this paper, we further approximate $I_{\mathrm{out}}$ by the first two terms of the Taylor approximation, which is then named as the scaling term $z_{\mathrm{DC}}$. The scaling term $z_{\mathrm{DC}}$ is extended in \cite{varasteh2019swipt} to wideband OFDM signals with random CSs, as a function of CS input distribution. Specifically,  $z_{\mathrm{DC}}$ under OFDM is written as}
\begin{subequations}
\begin{align}
\label{eq_scaling_term0}
z_{\mathrm{DC}}=&\mathbb{E}_{x}\left\lbrace\mathcal{E}\left[ k_2y_{\mathrm{RF}}(t)^2+k_4y_{\mathrm{RF}}(t)^4\right]\right\rbrace\\\nonumber
=&\sum_{m=0}^{M-1}\sum_{n=0}^{K'-1}\frac{k_2}{M} \mathbb{E}_{x}\left\lbrace |y[n,~m]|^2\right\rbrace+\\\label{eq_scaling_term1}&
\sum_{m=0}^{M-1}\sum_{n=0}^{K'-1} \frac{3k_4}{4M}\left(\mathbb{E}_{x}\left\lbrace |y[n,~m]|^4\right\rbrace+\mathbb{E}_{x}\left\lbrace |\widetilde{ y}[n,~m]|^4\right\rbrace  \right),
\end{align}
\end{subequations}
where $k_2=0.024$ and $k_4=19.145$ \cite{clerckx2016waveform}. In \eqref{eq_scaling_term1}, denote $y(t)$ as the complex baseband signal of $y_{\mathrm{RF}}(t)$. Then, $y[n,~m]$ and $\widetilde{y}[n,~m]$ are the samples of  $y(t)$  taken at times $t={(K'm+n)}/{B}$ and ${(2K'm+2n+1)}/{(2B)}$, which collectively form a sequence of samples capturing the power of $y(t)$ (based on the sampling theorem and Parseval's theorem [34]).  It is noteworthy that \cite{varasteh2019swipt} does not account for the harvested power attributed by the CP part of OFDM signals when expressing \eqref{eq_scaling_term1} as a function of the input distribution, which we address in the following.

If sampling at $t={(K'm+n)}/{B}$, $y[n,~m]$ is given by
\begin{subequations}\label{eq_samples0}
\begin{align}
y[n,~m]=&y({(K'm+n)}/{B})\\
=&\sum_l a_l  \sum_{q} x_{\mathrm{CP}}[q]\mathrm{sinc}\left[{K'm+n}-q-l\right]+w[n,~m]   \\
=&\sum_l a_l   x_{\mathrm{CP}}[{K'm+n}-l] +w[n,~m] \\\label{eq_samples01}
=&\sum_l a_l x_{\mathrm{CP}}[n-l,~m]+w[n,~m],
\end{align}
\end{subequations}
where $x_{\mathrm{CP}}[n-l,~m]=x_{\mathrm{CP}}[K'+n-l,~m-1]$ for $n<l$. $w[n,~m]$ is the sampled noise at ${y}[n,~m]$.

Similarly, if sampling at $t=({2K'm+2n+1})/{2B}$, $\widetilde{y}[n,~m]$ is given by
\begin{subequations}\label{eq_samples}
\begin{align}
\widetilde{y}[n,~m]=&y(({2K'm+2n+1})/{2B})\\
=&\sum_l a_l  \sum_{q}x_{\mathrm{CP}}[q]\text{sinc}\left[\frac{2K'm+2n+1}{2}-q-\tau_l\right]+w[n,~m]   \\\nonumber
{=}& \sum_j x_{\mathrm{CP}}[K'm+n-j]\sum_l a_l \text{sinc}\left[j+\frac{1}{2}-l\right]+w[n,~m] \\
&~~({j=K'm+n-q}) 
\\\label{eq_samples1}
\overset{\vartriangle}{=}&\sum_j \widetilde{a}_j x_{\mathrm{CP}}[n-j,~m]+\widetilde{w}[n,~m],\\\label{eq_samples1_non2}
\mbox{where}~\widetilde{a}_j=&\begin{cases}
    \sum_l a_l \text{sinc}\left[j+\frac{1}{2}-l\right],&~0\leq j\leq L-1;\\
    0~(\text{approximate}),&~j> L,
\end{cases}
\end{align}
\end{subequations}
and $\widetilde{w}[n,~m]$ is the sampled noise at $\widetilde{y}[n,~m]$. 
\begin{nrem}
{$\widetilde{y}[n,~m]$ in \eqref{eq_samples1} is similar  to ${y}[n,~m]$ in \eqref{eq_samples01}, with $a_l$ substituted by $\widetilde{a}_l$. Hence, most of the future analysis related with ${y}[n,~m]$ can be adopted to $\widetilde{y}[n,~m]$ directly after changing the coefficient.}
\end{nrem}
Given any $m$, the expression of ${y}[n,~m]$ in   \eqref{eq_samples01} (or $\widetilde{y}[n,~m]$ in  \eqref{eq_samples1}) are classified into two parts: (i) the CP duration for $0\leq n \leq K_G-1$ over which the received sub-pulses involves interaction between the $m^{\mathrm{th}}$ OFDM symbol and the $(m-1)^{\mathrm{th}}$  OFDM symbol; (ii) the  data duration for $K_G \leq n \leq K'-1$ where there is no inter-symbol interference. Hence, the final harvested power in  \eqref{eq_scaling_term1} also needs consideration from these two parts.

In the following, we first consider the sub-pulses' power over the $m^{\mathrm{th}}$  data duration ($K_G \leq n \leq K'-1$) and then move to the sub-pulses' power over the $m^{\mathrm{th}}$ CP duration ($0\leq n \leq K_G-1$). 

\subsubsection{Power of data duration}
starting from the data duration ($K_G \leq n \leq K'-1$), we take $y[n,~m]$  as an example first. $y[n,~m]$ for $K_G \leq n \leq K'-1$ can be re-expressed as
\begin{subequations}\label{eq_y_[n,m]}
\begin{align}
y[n,~m]=&\text{IDFT}\left\lbrace\mathbf{H}_\mathrm{P} \mathbf{x}_m \right\rbrace_{(n)}+w[n,~m]
\\=&{\mathbf{a}^{\mathrm{R}}_n}^T\mathbf{\overline{x}}_m+j{\mathbf{a}^{\mathrm{I}}_n}^T\mathbf{\overline{x}}_m+w[n,~m],\\
 \mbox{with}~
 \mathbf{H}_\mathrm{P}=&\text{diag}\left([h_{\mathrm{P,~0}},~\cdots,~h_{\mathrm{P,~K-1}}]\right),\\
h_{\mathrm{P},~k}=&\text{IDFT}\left\lbrace a_l \right\rbrace_{(k)},~\text{the frequency-domain channel,}\\{\mathbf{a}^{\mathrm{R}}_n}=&[\mathfrak{R}\left\lbrace {\mathbf{f}_{n}\mathbf{H}_{\mathrm{P}}} \right\rbrace,~-\mathfrak{I}\left\lbrace {\mathbf{f}_{n}\mathbf{H}_{\mathrm{P}}} \right\rbrace]^T,\\
 {\mathbf{a}^{\mathrm{I}}_n}=&[\mathfrak{I}\left\lbrace {\mathbf{f}_{n}\mathbf{H}_{\mathrm{P}}} \right\rbrace,~\mathfrak{R}\left\lbrace {\mathbf{f}_{n}\mathbf{H}_{\mathrm{P}}}
 \right\rbrace]^T,
\end{align}   
\end{subequations}

Similarly, for  $\widetilde{y}[n,~m]$, we have
\begin{subequations}\label{eq_y_[n,m]_tilde}
\begin{align}
\widetilde{y}[n,~m]=&\mbox{$\widetilde{\mathbf{a}}^{\mathrm{R}}_n$}^T\mathbf{\overline{x}}_m+j\mbox{$\widetilde{\mathbf{a}}^{\mathrm{I}}_n$}^T\mathbf{\overline{x}}_m+\widetilde{w}[n,~m],\\
 \mbox{with}~
 \widetilde{\mathbf{H}}_\mathrm{P}{=}&\text{diag}\left(\left[\widetilde{h}_{\mathrm{P,~0}},~\cdots,~\widetilde{h}_{\mathrm{P,~K-1}}\right]\right),\\
\widetilde{h}_{\mathrm{P},~k}=&\text{IDFT}\left\lbrace \widetilde{a}_l \right\rbrace_{(k)},\\
{\widetilde{ \mathbf{a}}^{\mathrm{R}}_n}=&\left[\mathfrak{R}\left\lbrace {\mathbf{f}_{n}\widetilde{\mathbf{H}}_{\mathrm{P}}} \right\rbrace,~-\mathfrak{I}\left\lbrace {\mathbf{f}_{n}\widetilde{\mathbf{H}}_{\mathrm{P}}} \right\rbrace\right]^T,\\
{\widetilde{ \mathbf{a}}^{\mathrm{I}}_n}=&\left[\mathfrak{I}\left\lbrace {\mathbf{f}_{n}\widetilde{\mathbf{H}}_{\mathrm{P}}} \right\rbrace,~\mathfrak{R}\left\lbrace {\mathbf{f}_{n}\widetilde{\mathbf{H}}_{\mathrm{P}}}
 \right\rbrace\right]^T.
\end{align}   
\end{subequations}

On this basis, the second-order and the fourth-order expectations of $y[n,~m]$ in \eqref{eq_scaling_term1}, i.e., $\mathbb{E}\left\lbrace |y[n,~m]|^2\right\rbrace$ and $\mathbb{E}\left\lbrace |y[n,~m]|^4\right\rbrace$, are given in  \eqref{eq_mean_OFDM} and \eqref{eq_4th_mean_OFDM} respectively (Details in Appendix \ref{appen1}). Following the same way, we also give the fourth-order expectations of $\widetilde{y}[n,~m]$, i.e.,  $\mathbb{E}\left\lbrace |\widetilde{y}[n,~m]|^4\right\rbrace$, in   \eqref{eq_4th_mean2_OFDM}. 
\begin{subequations}\label{eq_mean_OFDM}
\begin{align}
\mathbb{E}\left\lbrace |y[n,~m]|^2\right\rbrace=&\mathrm{Tr}\left(\mathbf{A}_n\mathbf{P} \right)+\sigma_{\mathrm{n_P}}^2
\overset{\vartriangle}{=}z^{\mathrm{data},~2}_{\mathrm{DC},~n}\left(\mathbf{P},~\mathbf{U}\right),\\\nonumber\nonumber\label{eq_4th_mean_OFDM}
\mathbb{E}\left\lbrace |y[n,~m]|^4\right\rbrace
=&\mathrm{Tr}^2\left(\mathbf{A}_{n}\mathbf{P}\right)+6\sigma_{\mathrm{n_P}}^2\mathrm{Tr}\left(\mathbf{A}_{n}\mathbf{P}\right)+2\mathrm{Tr}\left(\mathbf{A}_{n}\mathbf{P}\mathbf{A}_{n}\mathbf{P}\right)\\
&-2\mathrm{Tr}\left(\mathbf{A}_{n}\mathbf{U}\mathbf{A}_{n}\mathbf{U}\right)+3\sigma_{\mathrm{n_P}}^4\\\overset{\vartriangle}{=}&z^{\mathrm{data},~4}_{\mathrm{DC},~n}\left(\mathbf{P},~\mathbf{U}\right),\\\nonumber\nonumber\label{eq_4th_mean2_OFDM}
\mathbb{E}\left\lbrace |\widetilde{y}[n,~m]|^4\right\rbrace
=&\mathrm{Tr}^2\left(\widetilde{\mathbf{A}}_{n}\mathbf{P}\right)+6\sigma_{\mathrm{n_P}}^2\mathrm{Tr}\left(\mathbf{A}_{n}\mathbf{P}\right)+2\mathrm{Tr}\left(\widetilde{\mathbf{A}}_{n}\mathbf{P}\widetilde{\mathbf{A}}_{n}\mathbf{P}\right)\\
&-
2\mathrm{Tr}\left(\widetilde{\mathbf{A}}_{n}\mathbf{U}\widetilde{\mathbf{A}}_{n}\mathbf{U}\right)+3\sigma_{\mathrm{n_P}}^4\\
\overset{\vartriangle}{=}&\widetilde{z}^{\mathrm{data},~4}_{\mathrm{DC},~n}\left(\mathbf{P},~\mathbf{U}\right),\\
\mbox{with}~
\mathbf{A}_{n}=&{\mathbf{a}^{\mathrm{R}}_n}{\mathbf{a}^{\mathrm{R}}_n}^T+{\mathbf{a}^{\mathrm{I}}_n}{\mathbf{a}^{\mathrm{I}}_n}^T,\\
\widetilde{\mathbf{A}}_{n}=&\mbox{$\widetilde{\mathbf{a}}^{\mathrm{R}}_n$}\mbox{$\widetilde{\mathbf{a}}^{\mathrm{R}}_n$}^T+\mbox{$\widetilde{\mathbf{a}}^{\mathrm{I}}_n$}\mbox{$\widetilde{\mathbf{a}}^{\mathrm{I}}_n$}^T.
\end{align}
\end{subequations}

\subsubsection{Power of  CP duration}
based on   \eqref{eq_samples01}, $y[n,~m]$ for $0~\leq n ~\leq K_G-1$ can be re-expressed as
\begin{subequations}\label{eq_y_[n,m]_CP}
\begin{align}
\nonumber y[n,~m]
=&\sum^{n}_{l=0}a_lx_{\mathrm{CP}}[n-l,~m]+\sum^{K_G-1}_{l=n+1}a_lx_{\mathrm{CP}}[K'+n-l,~m-1]\\
&+w[n,~m]\\\nonumber
=&\sum^{n}_{l=0}a_lx[K-K_G+n-l,~m]\\
&+\sum^{K_G-1}_{l=n+1}a_lx[K+n-l,~m-1]+w[n,~m]\\\nonumber
=&\sum^{K-K_G+n}_{p=K-K_G}a_{K-K_G+n-p}x[p,~m]\\
&+\sum^{K-1}_{p=K-K_G+n+1}a_{K+n-p}x[p,~m-1]+w[n,~m]\\
=&{\mathbf{b}_{1,~n}}^T\mathbf{x}_m+{\mathbf{b}_{2,~n}}^T\mathbf{{x}}_{m-1}+w[n,~m],\\\nonumber
=&\mbox{$\mathbf{b}^{\mathrm{R}}_{1,~n}$}^T\mathbf{\overline{x}}_m+\mbox{$\mathbf{b}^{\mathrm{R}}_{2,~n}$}^T\mathbf{\overline{x}}_{m-1}\\
&+j\left(\mbox{$\mathbf{b}^{\mathrm{I}}_{1,~n}$}^T\mathbf{\overline{x}}_m+\mbox{$\mathbf{b}^{\mathrm{I}}_{2,~n}$}^T\mathbf{\overline{x}}_{m-1} \right) +w[n,~m],\\\label{eq_com_b}
\mbox{with}~
{\mathbf{b}_{i,~n}  =}&\text{DFT}{\left\lbrace \mathbf{a}_{\mathrm{CP}i,~n}\right\rbrace,~\text{for }i=1,~2},\\
 {a}_{\mathrm{CP}1,~n,~k}=&\begin{cases}
a_{K-K_G+n-k},~&K-K_G\leq k\leq K-K_G+n,\\
0,~&\text{otherwise},
\end{cases}\\\label{eq_com_b2}
 {a}_{\mathrm{CP}2,~n,~k}=&\begin{cases}
a_{K+n-k},~&K-K_G+n-1~\leq k~\leq K-1,\\
0,~&\text{otherwise},
\end{cases}\\
\mbox{$\mathbf{b}^{\mathrm{R}}_{i,~n}$}=&\left[\mathfrak{R}\left\lbrace {\mathbf{b}_{i,~n}} ^T\right\rbrace,~-\mathfrak{I}\left\lbrace {\mathbf{b}_{i,~n}}^T\right\rbrace\right]^T,\\\label{eq_com_b22}
\mbox{$\mathbf{b}^{\mathrm{I}}_{i,~n}$}=&\left[\mathfrak{I}\left\lbrace {\mathbf{b}_{i,~n}} ^T\right\rbrace,~\mathfrak{R}\left\lbrace {\mathbf{b}_{i,~n}}^T\right\rbrace\right]^T.
\end{align}
\end{subequations}

Similarly, $\widetilde{y}[n,~m]$ ($0~\leq n ~\leq K_G-1$) can be re-written as
\begin{align}\nonumber \label{eq_y_[n,m]_CP_tilde}
\widetilde{y}[n,~m]
=&\mbox{$\widetilde{\mathbf{b}}^{\mathrm{R}}_{1,~n}$}^T\mathbf{\overline{x}}_m+\mbox{$\widetilde{\mathbf{b}}^{\mathrm{R}}_{2,~n}$}^T\mathbf{\overline{x}}_{m-1}+j\left(\mbox{$\widetilde{\mathbf{b}}^{\mathrm{I}}_{1,~n}$}^T\mathbf{\overline{x}}_m+\mbox{$\widetilde{\mathbf{b}}^{\mathrm{I}}_{2,~n}$}^T\mathbf{\overline{x}}_{m-1} \right) \\
&+\widetilde{w}[n,~m], 
\end{align}
{where the corresponding expressions for $ {\widetilde{\mathbf{b}}^{\mathrm{R}}_{i,~n}}$ and $\widetilde{\mathbf{b}}^{\mathrm{I}}_{i,~n}$ $(i=1,~2)$  are obtained by replacing $a_l$ with $\widetilde{a}_l$ respectively in \eqref{eq_com_b}-\eqref{eq_com_b22}.}

Consequently, the second-order and the fourth-order expectations of the CP duration sub-pulses are (Details in Appendix \ref{appen2})
\begin{subequations}\label{eq_mean_CP}
   \begin{align}
\mathbb{E}\left\lbrace |y[n,~m]|^2\right\rbrace=&\mathrm{Tr}\left(\mathbf{B}_n\mathbf{P}+2\mathbf{D}_n\mathbf{U}\right)+\sigma_{\mathrm{n_P}}^2\overset{\vartriangle}{=}z^{\mathrm{CP},~2}_{\mathrm{DC},~n}\left(\mathbf{P},~\mathbf{U}\right),\\\nonumber
\mathbb{E}\left\lbrace |y[n,~m]|^4\right\rbrace
=&\mathrm{Tr}^2\left({\mathbf{B}}_n\mathbf{P}+2\mathbf{D}_n\mathbf{U}\right)  +6\sigma_{\mathrm{n_P}}^2\mathrm{Tr}\left(\mathbf{B}_n\mathbf{P}+2\mathbf{D}_n\mathbf{U}\right)\\\nonumber
&+2\mathrm{Tr}\left(\mathbf{B}_{1,~n}\mathbf{P}\mathbf{B}_{1,~n}\mathbf{P}+\mathbf{B}_{2,~n}\mathbf{P}\mathbf{B}_{2,~n}\mathbf{P}\right)\\\nonumber
&+4\mathrm{Tr}\left(\mathbf{D}_n\mathbf{P}\mathbf{D}_{n}^T\mathbf{P}+\mathbf{D}_{n}\mathbf{U}\mathbf{D}_{n}\mathbf{U}\right)\\\nonumber
&+4\mathrm{Tr}\left(\mathbf{B}_{1,~n}\mathbf{U}\mathbf{B}_{2,~n}\mathbf{U}+\mathbf{B}_{1,~n}\mathbf{P}\mathbf{D}_{n}\mathbf{U}\right)\\\nonumber
&+4\mathrm{Tr}\left(\mathbf{B}_{2,~n}\mathbf{P}\mathbf{D}^T_{n}\mathbf{U}+\mathbf{B}_{1,~n}\mathbf{U}\mathbf{D}^T_{n}\mathbf{P}\right)\\\nonumber
&+4\mathrm{Tr}\left(\mathbf{B}_{2,~n}\mathbf{U}\mathbf{D}_{n}\mathbf{P}\right)-2\mathrm{Tr}\left( \mathbf{E}_{n}\mathbf{U}\mathbf{E}_{n}^T\mathbf{U}\right)+3\sigma_{\mathrm{n_P}}^4\\\label{eq_mean4_CP}
\overset{\vartriangle}{=}&z^{\mathrm{CP},~4}_{\mathrm{DC},~n}\left(\mathbf{P},~\mathbf{U}\right),\\
\nonumber
\mathbb{E}\left\lbrace |\widetilde{y}[n,~m]|^4\right\rbrace
=&\mathrm{Tr}^2\left(\widetilde{\mathbf{B}}_n\mathbf{P}+2\widetilde{\mathbf{D}}_n\mathbf{U}\right)  +6\sigma_{\mathrm{n_P}}^2\mathrm{Tr}\left(\widetilde{\mathbf{B}}_n\mathbf{P}+2\widetilde{\mathbf{D}}_n\mathbf{U}\right)\\\nonumber
&+2\mathrm{Tr}\left(\widetilde{\mathbf{B}}_{1,~n}\mathbf{P}\widetilde{\mathbf{B}}_{1,~n}\mathbf{P}+\widetilde{\mathbf{B}}_{2,~n}\mathbf{P}\widetilde{\mathbf{B}}_{2,~n}\mathbf{P}\right)
\\\nonumber
&+4\mathrm{Tr}\left(\widetilde{\mathbf{D}}_n\mathbf{P}\widetilde{\mathbf{D}}_{n}^T\mathbf{P}+\widetilde{\mathbf{D}}_{n}\mathbf{U}\widetilde{\mathbf{D}}_{n}\mathbf{U}+\widetilde{\mathbf{B}}_{1,~n}\mathbf{U}\widetilde{\mathbf{B}}_{2,~n}\mathbf{U}\right)\\\nonumber
&+4\mathrm{Tr}\left(\widetilde{\mathbf{B}}_{1,~n}\mathbf{P}\widetilde{\mathbf{D}}_{n}\mathbf{U}+\widetilde{\mathbf{B}}_{2,~n}\mathbf{P}\widetilde{\mathbf{D}}^T_{n}\mathbf{U}\right)\\\nonumber
&+4\mathrm{Tr}\left(\widetilde{\mathbf{B}}_{1,~n}\mathbf{U}\widetilde{\mathbf{D}}^T_{n}\mathbf{P}+\widetilde{\mathbf{B}}_{2,~n}\mathbf{U}\widetilde{\mathbf{D}}_{n}\mathbf{P}\right)\\\nonumber
&-2\mathrm{Tr}\left( \widetilde{\mathbf{E}}_{n}\mathbf{U}\widetilde{\mathbf{E}}_{n}^T\mathbf{U}\right)+3\sigma_{\mathrm{n_P}}^4\\\label{eq_4th_CP2}
\overset{\vartriangle}{=}&\widetilde{z}^{\mathrm{CP},~4}_{\mathrm{DC},~n}\left(\mathbf{P},~\mathbf{U}\right),\\
\mbox{with}~\mathbf{B}_n=&\mathbf{B}_{1,~n}+\mathbf{B}_{2,~n},~\widetilde{\mathbf{B}}_n=\widetilde{\mathbf{B}}_{1,~n}+\widetilde{\mathbf{B}}_{2,~n},
\\
{\mathbf{B}_{i,~n}=}&{\mbox{$\mathbf{b}^\mathrm{R}_{i,~n}$}\mbox{$\mathbf{b}^\mathrm{R}_{i,~n}$}^T+\mbox{$\mathbf{b}^\mathrm{I}_{i,~n}$}\mbox{$\mathbf{b}^\mathrm{I}_{i,~n}$}^T,~\text{for }i=1,~2,}\\
{\widetilde{\mathbf{B}}_{i,~n}=}&{\mbox{$\widetilde{\mathbf{b}}^\mathrm{R}_{i,~n}$}\mbox{$\widetilde{\mathbf{b}}^\mathrm{R}_{i,~n}$}^T+\mbox{$\widetilde{\mathbf{b}}^\mathrm{I}_{i,~n}$}\mbox{$\widetilde{\mathbf{b}}^\mathrm{I}_{i,~n}$}^T,~\text{for }i=1,~2,}\\
\mathbf{D}_n=&\mbox{$\mathbf{b}^\mathrm{R}_{1,~n}$}\mbox{$\mathbf{b}^\mathrm{R}_{2,~n}$}^T+\mbox{$\mathbf{b}^\mathrm{I}_{1,~n}$}\mbox{$\mathbf{b}^\mathrm{I}_{2,~n}$}^T,\\
\widetilde{\mathbf{D}}_n=&\mbox{$\widetilde{\mathbf{b}}^\mathrm{R}_{1,~n}$}\mbox{$\widetilde{\mathbf{b}}^\mathrm{R}_{2,~n}$}^T+\mbox{$\widetilde{\mathbf{b}}^\mathrm{I}_{1,~n}$}\mbox{$\widetilde{\mathbf{b}}^\mathrm{I}_{2,~n}$}^T\\
\mathbf{E}_n=&{\mathbf{B}}_{1,~n}+\mathbf{D}_n+\mathbf{B}_{2,~n}^T+\mathbf{D}_n^T,\\
\widetilde{\mathbf{E}}_n=&\widetilde{\mathbf{B}}_{1,~n}+\widetilde{\mathbf{D}}_n+\widetilde{\mathbf{B}}_{2,~n}^T+\widetilde{\mathbf{D}}_n^T.
\end{align} 
\end{subequations}

\subsubsection{Total power}
After re-organization, the scaling term of the generated power of the $m^{\mathrm{th}}$ OFDM symbol as a function of the input distribution is given as 
\begin{align}
\nonumber&z_{\mathrm{DC}}\left(\mathbf{P},~\mathbf{U}\right)\\\nonumber\label{eq_z_DC_final}
=&\sum_{n=0}^{K_G-1} k_2 z^{\mathrm{CP},~2}_{\mathrm{DC},~n}\left(\mathbf{P},~\mathbf{U}\right)
+\frac{3k_4}{4}\left(z^{\mathrm{CP},~4}_{\mathrm{DC},~n}\left(\mathbf{P},~\mathbf{U}\right)+\widetilde{z}^{\mathrm{CP},~4}_{\mathrm{DC},~n}\left(\mathbf{P},~\mathbf{U}\right)\right)\\
&+
\sum_{n=K_G}^{K'-1} k_2 z^{\mathrm{data},~2}_{\mathrm{DC},~n}\left(\mathbf{P},~\mathbf{U}\right)+\frac{3k_4}{4}\left(z^{\mathrm{data},~4}_{\mathrm{DC},~n}\left(\mathbf{P},~\mathbf{U}\right)+\widetilde{z}^{\mathrm{data},~4}_{\mathrm{DC},~n}\left(\mathbf{P},~\mathbf{U}\right)\right).
\end{align}
\begin{nrem}\label{remark_WPT}
    \textcolor{blue}{The powering metric in \eqref{eq_z_DC_final} results in an asymmetric design between the real and imaginary parts of the CS input distribution. This is visualized from the received samples in \eqref{eq_y_[n,m]}, \eqref{eq_y_[n,m]_tilde}, \eqref{eq_y_[n,m]_CP} and \eqref{eq_y_[n,m]_CP_tilde} where the real and imaginary parts of the frequency-domain CS, $\overline{\mathbf{x}}_m$,  experiences different powering channels. Hence, optimizing the expected delivered power from the received samples, as outlined in \eqref{eq_z_DC_final}, yields an asymmetric power allocation between the real and imaginary parts of the CSs. Similar observations are in \cite{varasteh2019swipt}, where CSs with asymmetric input distribution outperform those with symmetric input distribution w.r.t the powering functionality.}
\end{nrem}

\subsection{Metric for Sensing}
\label{section_radar}
In terms of sensing, we consider a point-target sensing scenario aiming at the range-velocity bin estimation. The sensing's performance is guaranteed by restricting its aISPLD in terms of the ambiguity property. Indeed, according to \cite{2022ISAC}, the aISPLD is shown to be in proportion to the upper bound of the average FAP, i.e., the probability of incorrectly estimating the target's range-velocity bin using the maximum likelihood estimator in OFDM ISAC systems in low SNR. On this basis, we express our sensing metric,  $\overline{\mathrm{UB}}_{\mathrm{FAP}} $, as
\begin{subequations}\label{eq_aISPLD}
 \begin{align}\nonumber\label{eq_aISPLD1}
\overline{\mathrm{UB}}_{\mathrm{FAP}}(\mathbf{p},~\boldsymbol{\mu})=& -M(K_GM-1)\mathbf{1}^{T}\mathbf{{p}}\\
&+{\sum_{(r,~v)\neq (0,~0)}\sqrt{g_{r,v}(\mathbf{{p}})+g_2(\boldsymbol{\mu})}},\\\label{eq_average_UB_function_vector_gnv}
\mbox{with}~
g_{r,v}(\mathbf{{p}})
=&M^2\delta\left(v\right)|\left(\mathbf{1}_{2}^T\otimes\mathbf{f}_{r}\right)\mathbf{{p}}|^2+2M\Vert\mathbf{{p}}\Vert^2,\\
g_2(\boldsymbol{\mu})=&-2M\sum_{k=0}^{2K-1}|{\mu}_k|^4,
\end{align}   
\end{subequations}
where  $(r,~v)$ represents the range-velocity bin candidate for the OFDM radar system, with $r=0, ~\cdots,~K_G-1$ and $v=-M/2,~\cdots,~ M/2-1$. Assuming a target at $(r,~v)=(0,~0)$, the first sum term in   \eqref{eq_aISPLD1} represents the average peak lobe at  $(r,~v)=(0,~0)$ and the second sum term in  \eqref{eq_aISPLD1} represents the upper bound of the average sum of side lobes where $(r,~v)\neq (0,~0)$.

\begin{nrem} Given a transmit power constraint, i.e., a fixed $\mathbf{1}^{T}\mathbf{{p}}$, the aISPLD metric in \eqref{eq_aISPLD} is minimized by maximizing the symbol mean at each subcarrier, i.e.,  $\boldsymbol{\mu}=\mathbf{p}$, while the optimal $\mathbf{p}$ is to be uniform across sub-carriers from \cite{2022ISAC}. Intuitively, sensing prefers deterministic components that are represented by the symbol mean of the input distribution, in contrast with communications. In other words, the randomness of the CS with Gaussian inputs degrades sensing. 
\end{nrem}

\begin{nrem}\label{remark_sensing}
    \textcolor{blue}{For the special case of $\boldsymbol{\mu}=\mathbf{p}$ (no communications) under a fixed transmit power constraint (fixed $\mathbf{1}^{T}\mathbf{{p}}$), the aISPLD metric in \eqref{eq_aISPLD} depends only on the terms $|\left(\mathbf{1}_{2}^T\otimes\mathbf{f}_{r}\right)\mathbf{{p}}|^2$ for $r=0,~\cdots,~K_G-1$, or equivalently, depends only on the combined terms $p_{k}+p_{k+K}$ for $k=0,~\cdots,~K-1$. This result indicates that the sensing metric is independent of the power ratio between the real (i.e., $p_{k}$) and imaginary (i.e., $p_{k+K}$) parts of OFDM CSs, as long as the total power allocation to sub-carrier $k$ for $k=0,~\cdots,~K-1$ has been fixed. In this context, the asymmetric design between the real and imaginary parts does not bring much benefit to sensing performance,  which is different from the powering metric as mentioned in Remark \ref{remark_WPT}. This observation indicates that the sensing performance is more influenced by the power allocation trade-off between $\boldsymbol{\mu}$ and $\boldsymbol{\sigma}$ (the random-deterministic trade-off) than the power allocation trade-off between the real and imaginary components. }
\end{nrem}

Similar to the powering metric expression in \eqref{eq_z_DC_final}, we re-formulate the sensing metric $\overline{\mathrm{UB}}_{\mathrm{FAP}}(\mathbf{p},~\boldsymbol{\mu})$ in \eqref{eq_aISPLD} as a function of $\mathbf{P}$ and $\mathbf{U}$ as follows
\begin{subequations}\label{eq_SPCA_UB}
 \begin{align}
\nonumber
\widetilde{\mathrm{UB}}_{\mathrm{FAP}}(\mathbf{P},~\mathbf{U})
= &-M(K_GM-1)\mathrm{Tr}\left(\mathbf{P}\right)\\&
+{\sum_{(r,~v)\neq (0,~0)}\sqrt{\widetilde{g}_{r,v}\left(\mathbf{P}\right)+\widetilde{g}_2\left(\mathbf{U}\right)}},\\
\mbox{with}~
\widetilde{g}_{r,v}\left(\mathbf{P}\right)=& \left({\mathbf{f}^{\mathrm{R}}_{r,v}}\text{diag}\left( \mathbf{P}\right)\right)^2+ \left( {\mathbf{f}^{\mathrm{I}}_{r,v}}\text{diag}\left( \mathbf{P}\right)\right)^2+2 M\Vert\text{diag}\left( \mathbf{P}\right)\Vert^2,\\\label{eq_average_UB_function_vector}
\widetilde{g}_2( \mathbf{U})=&-2M\|\text{diag}\left( \mathbf{U}\right)\|^2,\\
\mathbf{f}^{\mathrm{R/I}}_{r,v}=&M\delta\left(v \right)\mathfrak{R/I}\left\lbrace \mathbf{1}_{2}^T\otimes\mathbf{f}_{n} \right\rbrace.
\end{align}   
\end{subequations}

\subsection{Metric for Communications}
\label{section_communication}
For the communication system, \textcolor{blue}{similarly to the powering channel, we assume a static communication channel, whose CSI has been perfectly estimated in advance through pilot signals.} In this context, the complex received signal at the $k^{\mathrm{th}}$ subcarrier is expressed as
\begin{align}
\label{eq_y_communication_signal}
Y_{\mathrm{C}}[k,~m]=& h_{\mathrm{C},~k}X[k,~m]+w_{\mathrm{C}}[k,~m],
\end{align}
where $h_{\mathrm{C},~k}$ is the complex gain of the communication channel of the $k^{\mathrm{th}}$ subcarrier. $w_{\mathrm{C}}[k,~m]$ is the AWGN at the communication receiver with its noise power spectrum density being $\sigma_{\mathrm{n_C}}^2$.

Given that the variance of the real and imaginary part of $X[k,~m]$ are ${\sigma^{\mathrm{R}}_{k}}^2$ and ${\sigma^{\mathrm{I}}_{k}}^2$ respectively, the average communication achievable rate in OFDM can be written as\cite{varasteh2019swipt}
 \begin{align}
\nonumber\label{eq_MI_communication_scaler}
R\left(\left\lbrace \sigma^{\mathrm{R}}_{k},~\sigma^{\mathrm{I}}_{k} \right\rbrace\right)=&\frac{1}{2KM}\sum_{k,~m}\log~\left(1+{2K|h_{\mathrm{C},~k}|^2}{ \sigma^{\mathrm{R}}_{k}}^2/(B\sigma^2_{\mathrm{n_C}})\right)
\\&+\frac{1}{2KM}\sum_{k,~m}\log~\left(1+{2K|h_{\mathrm{C},~k}|^2}{ \sigma^{\mathrm{I}}_{k}}^2/(B\sigma^2_{\mathrm{n_C}})\right).
\end{align}  

\eqref{eq_MI_communication_scaler} can be re-formulated in the vector form as
\begin{subequations}
 \label{eq_MI_communication}\begin{align}
R(\boldsymbol{\sigma})=&\frac{1}{2K}\log~\det \left[\mathbf{I}_{2K}+\frac{2K\mathbf{H}_{\mathrm{C}}\mathrm{diag}(\boldsymbol{\sigma})}{B\sigma^2_{\mathrm{n_C}}}\right],\\
\mbox{with}~\mathbf{H}_{\mathrm{C}}=&\text{diag}\left(\mathbf{1}_2\otimes{\left[{h}_{\mathrm{C},~0},~\cdots,~{h}_{\mathrm{C},~K-1}^T\right]}\right).
\end{align}   
\end{subequations}

\textcolor{blue}{
 For convenient reference, we summarize the mathematical notations of important coefficients involved in different functionality metrics from \eqref{eq_samples0} to \eqref{eq_MI_communication} in Table \ref{tab1}.}
\begin{table*}\color{blue}
     \centering
     \begin{tabular}{|p{2.8cm}|p{14.5 cm}|}
    \hline
    Notation  &  Definition/Intuition \\\hline
    $a_l$&The complex gain of the $l^\mathrm{th}$ tap in the WPT channel for $y[n,~m]$ in \eqref{eq_samples0}\\\hline
    $\widetilde{a}_l$&The (analytical) complex gain of the $l^\mathrm{th}$ tap in the WPT channel for $\widetilde{y}[n,~m]$ in \eqref{eq_samples}\\\hline
    $\mathbf{a}^{\mathrm{R}}_n$  ($\mathbf{a}^{\mathrm{I}}_n$)& The transformation coefficients from the frequency-domain $\overline{\mathbf{x}}_m$ to the real (imaginary) part of time-domain subpulse $y[n,~m]$ over the OFDM data duration, i.e., $n\geq K_G$, in the WPT channel in \eqref{eq_y_[n,m]}\\\hline
    ${\widetilde{ \mathbf{a}}^{\mathrm{R}}_n}$ (${\widetilde{ \mathbf{a}}^{\mathrm{I}}_n}$)&The transformation coefficients from the frequency-domain $\overline{\mathbf{x}}_m$ to the real (imaginary) part of time-domain subpulse $\widetilde{y}[n,~m]$  over the OFDM data duration, i.e., $n\geq K_G$, in the WPT channel in \eqref{eq_y_[n,m]_tilde}\\\hline
        $\mathbf{A}_{n}$ (${\widetilde{\mathbf{A}}_n}$)&The WPT coefficient matrix defined for $\mathbf{P}$ and $\mathbf{U}$ -- formed by $\mathbf{a}^{\mathrm{R}}_n$ and $\mathbf{a}^{\mathrm{I}}_n$ ($\widetilde{ \mathbf{a}}^{\mathrm{R}}_n$ and $\widetilde{ \mathbf{a}}^{\mathrm{I}}_n$)-- corresponding to the generated power from $y[n,~m]$ ($\widetilde{y}[n,~m]$)  over the OFDM data duration, i.e., $n\geq K_G$, in \eqref{eq_mean_OFDM} \\\hline
        $\mathbf{b}^{\mathrm{R}}_{i,n}$  ($\mathbf{b}^{\mathrm{I}}_{i,n}$)& The transformation coefficients from the frequency-domain $\overline{\mathbf{x}}_m,~i=1$ or $\overline{\mathbf{x}}_{m-1},~i=2$ to the real (imaginary) part of time-domain subpulse $y[n,~m]$ over the OFDM CP duration, i.e., $n< K_G$, in the WPT channel in \eqref{eq_y_[n,m]_CP}\\\hline
    $\widetilde{\mathbf{b}}^{\mathrm{R}}_{i,n}$  ($\widetilde{\mathbf{b}}^{\mathrm{I}}_{i,n}$)&The transformation coefficients from the frequency-domain $\overline{\mathbf{x}}_m,~i=1$ or $\overline{\mathbf{x}}_{m-1},~i=2$ to the real (imaginary) part of time-domain subpulse $\widetilde{y}[n,~m]$ over the OFDM CP duration, i.e., $n< K_G$, in the WPT channel in \eqref{eq_y_[n,m]_CP_tilde}\\\hline
        $\mathbf{B}_{i,~n}$ ($\widetilde{\mathbf{B}}_{i,~n}$), $\mathbf{D}_{n}$ ($\widetilde{\mathbf{D}}_{n}$), $\mathbf{E}_{n}$ ($\widetilde{\mathbf{E}}_{n}$)&The WPT coefficient matrices defined for $\mathbf{P}$ and $\mathbf{U}$ -- formed by $\mathbf{b}^{\mathrm{R}}_{i,n}$ and $\mathbf{b}^{\mathrm{I}}_{i,n}$ ($\widetilde{ \mathbf{b}}^{\mathrm{R}}_{i,n}$ and $\widetilde{ \mathbf{b}}^{\mathrm{I}}_{i,n}$)-- corresponding to the generated power from $y[n,~m]$ ($\widetilde{y}[n,~m]$)  over the OFDM CP duration, i.e., $n< K_G$, in \eqref{eq_mean_CP}\\\hline
        $\mathbf{f}^{\mathrm{R/I}}_{r,v}$& The coefficient vector in \eqref{eq_aISPLD} to formulate sidelobes of the 
        ambiguity function from $\mathbf{P}$ for OFDM radar \cite{2022ISAC}\\\hline 
        $h_{\mathrm{C},~k}$, $\mathbf{H}_{\mathrm{C}}$& The complex gain of the communication channel of the $k^{\mathrm{th}}$ subcarrier in \eqref{eq_y_communication_signal}-\eqref{eq_MI_communication}; $\mathbf{H}_{\mathrm{C}}$ is the matrix form\\\hline 
\end{tabular}
     \caption{Table of the important coefficients involved in the three functionality metrics. }
     \label{tab1}
 \end{table*}

\section{Optimal OFDM Waveform Design}
\label{section_opt}
In this section, we formulate the problem to optimize the performance region of the ISCAP system, i.e., to maximize the harvested power per symbol given constraints on the transmit power, the achievable rate (communication) and the aISPLD (sensing). Upon formulating the problem which is non-convex, we propose an ADMM-based algorithm to obtain a local optimal solution.

\subsection{Problem formulation}
The optimization problem that maximizes the harvested power per symbol given achievable rate constraint and the aISPLD constraint is formulated as 
\begin{maxi!}
        {{\mathbf{P}},~{\mathbf{U}},~\boldsymbol{\sigma}}{z_{\mathrm{DC}}({\mathbf{P}},~\mathbf{U}) \text{ in } \eqref{eq_z_DC_final},}{\label{eq_optimization_P1}}{\label{eq_optimization_P1_1}}
        \addConstraint{ \mathrm{Tr}\left({\mathbf{P}}\right)\leq P_{\max}}\label{eq_optimization_P1_2}
        \addConstraint{R(\boldsymbol{\sigma}) \geq C_{\min}}\label{eq_optimization_P1_3}
        \addConstraint{\widetilde{\mathrm{UB}}_{\mathrm{FAP}}(\mathbf{P},~\mathbf{U}) \leq S_{\max}}\label{eq_optimization_P1_4}       
        \addConstraint{\mathbf{U}\geq 0}\label{eq_optimization_P1_5} 
        \addConstraint{\mathrm{rank}\left(\mathbf{U}\right)= 1}\label{eq_optimization_P1_52}  
        \addConstraint{\boldsymbol{\sigma}\succeq 0}\label{eq_optimization_P1_6}
        \addConstraint{\mathbf{P}=\mathbf{U}+\text{diag}\left(\boldsymbol{\sigma}\right),}\label{eq_optimization_P1_7}
      \end{maxi!}
where $P_{\max}$ in \eqref{eq_optimization_P1_2} is the transmit power constraint at the transmitter, $C_{\min}$ in  \eqref{eq_optimization_P1_3} is the minimum rate constraint for communication, and $S_{\max}$ in   \eqref{eq_optimization_P1_4} is the maximal aISPLD constraint for sensing.  Constraint \eqref{eq_optimization_P1_5} and   \eqref{eq_optimization_P1_52} ensure  a solution for $\mathbf{U}=\boldsymbol{\mu}\boldsymbol{\mu}^T$.   \eqref{eq_optimization_P1_6} ensures  positive covariance and   \eqref{eq_optimization_P1_7} establishes the relationship between the variables as defined in \eqref{eq_define_U}-\eqref{eq_define_P}.

\subsection{Problem optimization}
The rank 1 constraint in problem \eqref{eq_optimization_P1} is relaxed first during the optimization and is handled at the final stage by   Gaussian randomization\footnote{\textcolor{blue}{Although Gaussian randomization is a heuristic method, it is widely recognized to provide a performance benchmark for solving rank 1 relaxation problems with satisfactory accuracy \cite{luo2010semidefinite,luo2010sdp}. As a substitute, we may also consider converting the rank 1 constraint into a penalty term in the objective function \cite{recht2010guaranteed}, adopting eigenvalue approximations rather than Gaussian randomization after the rank 1 relaxation \cite{luo2010semidefinite}, using sequential (rank) relaxation methods \cite{cao2017sequential}, or reformulating the problem by exploiting certain relaxed structural constraints \cite{yue2023joint}. While these substitute methods or low-complexity algorithms are beyond the scope of this paper, they present interesting future works.}} \cite{luo2010semidefinite}. On this basis, we solve the remaining optimization problem following an ADMM structure.  
      
The ADMM structure of problem \eqref{eq_optimization_P1}, relaxing the rank $1$ constraint in  \eqref{eq_optimization_P1_52}  is expressed as
\begin{mini!}
        {\mathbf{P},~{\mathbf{U}},~\boldsymbol{\sigma}}{
        -z_{\mathrm{DC}}({\mathbf{P}},~\mathbf{U}) +f_{\mathrm{I},~\mathbf{P},\mathbf{U}}\left(\mathbf{P},~\mathbf{U}\right)+f_{\mathrm{I},~\boldsymbol{\sigma},~\mathbf{U}}\left(\boldsymbol{\sigma},~\mathbf{U}\right),}{\label{eq_optimization_P2}}{\label{eq_optimization_P2_1}}
         \addConstraint{\mathbf{U}+\text{diag}\left(\boldsymbol{\sigma}\right)-\mathbf{P}=\mathbf{0},}\label{eq_optimization_P2_2}
      \end{mini!}
where $f_{\mathrm{I},~\mathbf{P},~\mathbf{U}}\left(\mathbf{P},~\mathbf{U}\right)$   is the indicator function of $\mathbf{P}$ and $\mathbf{U}$ corresponding to the constraint in \eqref{eq_optimization_P1_2} and \eqref{eq_optimization_P1_4}, and   $f_{\mathrm{I},~\boldsymbol{\sigma},~\mathbf{U}}\left(\boldsymbol{\sigma},~\mathbf{U}\right)$   is the indicator function of $\boldsymbol{\sigma}$ and $\mathbf{U}$ corresponding to the constraints in \eqref{eq_optimization_P1_3},~\eqref{eq_optimization_P1_4},~\eqref{eq_optimization_P1_5} and \eqref{eq_optimization_P1_6}, i.e., 
\begin{align}
f_{\mathrm{I},~\mathbf{P},~\mathbf{U}}\left(\mathbf{P},~\mathbf{U}\right):=&f_{\mathrm{I}}\left(  \mathrm{Tr}\left({\mathbf{P}}\right)\leq P_{\max}\right)+ f_{\mathrm{I}}\left( \widetilde{\mathrm{UB}}_{\mathrm{FAP}}(\mathbf{P},~\mathbf{U}) \leq S_{\max}\right),\\\nonumber
f_{\mathrm{I},~\boldsymbol{\sigma},~\mathbf{U}}\left(\boldsymbol{\sigma},~\mathbf{U}\right):=&f_{\mathrm{I}}\left(  R(\boldsymbol{\sigma}) \geq C_{\min} \right)+ f_{\mathrm{I}}\left( \widetilde{\mathrm{UB}}_{\mathrm{FAP}}(\mathbf{P},~\mathbf{U}) \leq S_{\max}\right)\\&+f_{\mathrm{I}}\left(  \mathbf{U}\geq 0\right)+ f_{\mathrm{I}}\left( \boldsymbol{\sigma}\succeq 0\right).
\end{align}

Following the ADMM structure in problem \eqref{eq_optimization_P2}, the variables of $\mathbf{P}$, $\mathbf{U}$ and $\boldsymbol{\sigma}$ at the $l^{\mathrm{th}}$ ADMM iteration are successively updated  as in \eqref{eq_ADMM_P}-\eqref{eq_ADMM_v}, where $\mathbf{V}^{(l)}$ is the alternating direction at the $l^{\mathrm{th}}$ iteration and $\rho$ is the updating step-size in ADMM.
\begin{figure*}
\begin{subequations}
 \begin{align}
\scriptstyle\mathbf{P}^{(l+1)}\label{eq_ADMM_P}
\scriptstyle:=&\scriptstyle\arg\min\left\lbrace{-z_{\mathrm{DC}}\left(\mathbf{P},~{\mathbf{U}}^{(l)}\right) +f_{\mathrm{I},~\mathbf{P},\mathbf{U}}\left(\mathbf{P},\mathbf{U}^{(l)}\right)+} {\frac {\rho}{2} \|{\mathbf{U}}^{(l)}-\mathbf{P}+\text{diag}\left({\boldsymbol{\sigma}}^{(l)}\right)-{\mathbf{V}}^{(l)}\|^2}\right\rbrace,\\\label{eq_ADMM_U}
\scriptstyle{\left\lbrace{\mathbf{U}}^{(l+1)},~{\boldsymbol{\sigma}}^{(l+1)}\right\rbrace}:=&\scriptstyle\arg\min \left\lbrace{-z_{\mathrm{DC}}\left({\mathbf{P}}^{(l+1)},~\mathbf{U}\right)+ f_{\mathrm{I},~\boldsymbol{\sigma},~\mathbf{U}}(\boldsymbol{\sigma},~\mathbf{U})+}{ \frac{\rho}{2}\|-{\mathbf{P}}^{(l+1)}+\mathbf{U}+\text{diag}\left({\boldsymbol{\sigma}}\right)-{\mathbf{V}}^{(l)}\|^2}\right\rbrace,\\\label{eq_ADMM_v}
\scriptstyle{{\mathbf{V}}^{(l+1)}:=}&\scriptstyle{{\mathbf{V}}^{(l)}+{\mathbf{P}}^{(l+1)}-{\mathbf{U}}^{(l+1)}-\text{diag}\left({\boldsymbol{\sigma}}^{(l+1)}\right)},
\end{align}   
\end{subequations}
\noindent\makebox[\linewidth]{\rule{\paperwidth}{0.4pt}}
\vspace*{-0.4cm}
\end{figure*}

Next, we solve the sub-problems in \eqref{eq_ADMM_P} and  \eqref{eq_ADMM_U} respectively to update the variables $\mathbf{P}$, $\mathbf{U}$ and $\boldsymbol{\sigma}$.

\subsubsection{Updating ${\mathbf{P}}^{(l+1)}$}
The sub-problem to solve  \eqref{eq_ADMM_P} is
\begin{mini!}
        {\mathbf{P}}{-z_{\mathrm{DC}}\left(\mathbf{P},~{\mathbf{U}}^{(l)}\right)+\frac{\rho}{2}\|-\mathbf{P}+{\mathbf{U}}^{(l)}+\text{diag}\left({\boldsymbol{\sigma}}^{(l)}\right)-{\mathbf{V}}^{(l)}\|^2}{\label{eq_optimization_P3}}{\label{eq_optimization_P3_1}}
        \addConstraint{ \mathrm{Tr}\left({\mathbf{P}}\right)\leq P_{\max}}\label{eq_optimization_P3_3}
\addConstraint{\widetilde{\mathrm{UB}}_{\mathrm{FAP}}\left(\mathbf{P},~{\mathbf{U}}^{(l)}\right) \leq S_{\max},}\label{eq_optimization_P3_2}   
      \end{mini!}
which is non-convex due to the objective  $-z_{\mathrm{DC}}\left(\mathbf{P},~{\mathbf{U}}^{(l)}\right)$ and the constraint \eqref{eq_optimization_P3_3}. However, $-z_{\mathrm{DC}}\left(\mathbf{P},~{\mathbf{U}}^{(l)}\right)$ in the objective function is concave with respect to $\mathbf{P}$, and can be handled by using SCA \footnote{\textcolor{blue}{The convergence of such an ADMM-SCP combination algorithm has been well established in \cite{scutari2016parallel,kumar2019distributed} and exploited in robust scenarios \cite{scutari2016parallel2,8972927}.}}. In SCA, the concave part of the objective function is successively substituted by a linear Taylor approximation operating at the point which is the local optimal of the previous round. Specifically, suppose $\mathbf{P}^{(t)}$ is the optimal point at the $t^{\mathrm{th}}$ SCA iteration, then, at the $ (t+1)^{\mathrm{th}}$ iteration,  $-z_{\mathrm{DC}}(\mathbf{P},{\mathbf{U}}^{(l)})$ in the objective function is substituted by
 \begin{align}
 \label{eq_SCP_P_SCA}
{z_{\mathrm{DC}}}^{(t+1)}\left(\mathbf{P},~{\mathbf{U}}^{(l)}\right)={z_{\mathrm{DC},~0}}^{(t)}\left(\mathbf{P}^{(t)},~{\mathbf{U}}^{(l)}\right)+\mathrm{Tr}\left({\mathbf{G}^{(t)}}^T\mathbf{P}\right),
 \end{align}
 where ${\mathbf{G}^{(t)}}$ is the first-order Taylor coefficient, expressed in  \eqref{eq_P_SCP_coeff}. ${z_{\mathrm{DC},~0}}^{(t)}\left(\mathbf{P}^{(t)},~{\mathbf{U}}^{(l)}\right)$ is the Taylor constant to be omitted.

\begin{figure*}
\begin{align}
\nonumber\scriptstyle{{\mathbf{G}^{(t)}}}=&\scriptstyle{\sum_{n=K_G}^{K'-1} k_2 \mathbf{A}_n^T +\frac{3k_4}{4}\left[ 2\mathrm{Tr}\left(\mathbf{A}_{n}\mathbf{P} ^{(t)}\right)\mathbf{A}_n^T+ 6\sigma_{\mathrm{n_P}}^2\mathbf{A}_n^T+4\mathbf{A}_n\mathbf{P}^{(t)}\mathbf{A}_n + 2\mathrm{Tr}\left(\widetilde{\mathbf{A}}_{n}\mathbf{P} ^{(t)}\right)\widetilde{\mathbf{A}}_n^T+ 6\sigma_{\mathrm{n_P}}^2\widetilde{\mathbf{A}}_n^T+4\widetilde{\mathbf{A}}_n\mathbf{P}^{(t)}\widetilde{\mathbf{A}}_n\right]}\\\nonumber
&\scriptstyle{ +\sum_{n=0}^{K_G-1} k_2 \mathbf{B}_n^T+\frac{3k_4}{4}\left[2\mathrm{Tr}\left(\mathbf{B}_{n}\mathbf{P}^{(t)}+2\mathbf{D}_{n}\mathbf{U}^{(l)}\right)\mathbf{B}_n^T+6\sigma_{\mathrm{n_P}}^2\mathbf{B}^T_n+2\mathrm{Tr}\left(\widetilde{\mathbf{B}}_{n}\mathbf{P}^{(t)}+2\widetilde{\mathbf{D}}_{n}\mathbf{U}^{(l)}\right)\widetilde{\mathbf{B}}_n^T+6\sigma_{\mathrm{n_P}}^2\widetilde{\mathbf{B}}^T_n\right]}\\\nonumber
&\scriptstyle{+\sum_{n=0}^{K_G-1}{3k_4}\left[\mathbf{B}_{1,~n}\left(\mathbf{P}^{(t)}\mathbf{B}_{1,~n}+\mathbf{U}^{(l)}\mathbf{D}_{n}^T\right)+\mathbf{D}^T_n\left(\mathbf{P}^{(t)}\mathbf{D}_{n}+\mathbf{U}^{(l)}\mathbf{B}_{2,~n}\right)+\mathbf{D}_n\left(\mathbf{P}^{(t)}\mathbf{D}_{n}^T+\mathbf{U}^{(l)}\mathbf{B}_{1,~n}\right)+\mathbf{B}_{2,~n}\left(\mathbf{P}^{(t)}\mathbf{B}_{2,~n}+\mathbf{U}^{(l)}\mathbf{D}_{n}\right)\right]}\\\label{eq_P_SCP_coeff}
&\scriptstyle{+\sum_{n=0}^{K_G-1}{3k_4}\left[\widetilde{\mathbf{B}}_{1,~n}\left(\mathbf{P}^{(t)}\widetilde{\mathbf{B}}_{1,~n} +\mathbf{U}^{(l)}\widetilde{\mathbf{D}}_{n}^T\right)+\widetilde{\mathbf{D}}^T_n\left(\mathbf{P}^{(t)}\widetilde{\mathbf{D}}_{n}+\mathbf{U}^{(l)}\widetilde{\mathbf{B}}_{2 ,~n}\right)+\widetilde{\mathbf{D}}_n\left(\mathbf{P}^{(t)}\widetilde{\mathbf{D}}_{n}^T+\mathbf{U}^{(l)}\widetilde{\mathbf{B}}_{1,~n}\right)+\widetilde{\mathbf{B}}_{2,~n}\left(\mathbf{P}^{(t)}\widetilde{\mathbf{B}}_{2 ,~n}+\mathbf{U}^{(l)}\widetilde{\mathbf{D}}_{n}\right)\right]}.
\end{align}
\noindent\makebox[\linewidth]{\rule{\paperwidth}{0.4pt}}
\vspace*{-0.4cm}
\end{figure*}

Similarly, the non-linear constraint in  \eqref{eq_optimization_P3_2} is also approximated by its linear upper bound in each iteration of SCA. Since the concavity of \eqref{eq_optimization_P3_2} comes from the square root function, we also approximate the square root function by using Taylor expansion  as 
\begin{subequations}
\begin{align}
\nonumber
&\widetilde{\mathrm{UB}}_{\mathrm{FAP}}\left(\mathbf{P},~{\mathbf{U}}^{(l)}\right)\\\nonumber
\leq & -M(K_GM-1) \mathrm{Tr}\left(\mathbf{P} \right)+\sum_{(r,~v)\neq(0,~0)}\frac{2\left(\widetilde{g}_{r,v}\left({\mathbf{P}}^{(t)}\right)+\widetilde{g}_2\left({\mathbf{U}}^{(l)}\right)\right)}{\alpha^{(l,t)}_{r,v}}+\\&\sum_{(r,~v)\neq(0,~0)}\frac{\left[\widetilde{g}_{r,v}\left({\mathbf{P}}\right)+\widetilde{g}_2\left(\mathbf{U}^{(l)}\right)-\widetilde{g}_{r,v}\left({\mathbf{P}^{(t)}}\right)-\widetilde{g}_2\left(\mathbf{U}^{(l)}\right)\right]}{\alpha^{(l,t)}_{r,v}}\\
=&\alpha^{(l,t)}-M(K_GM-1)  \mathrm{Tr}\left(\mathbf{P}\right)+\sum_{(r,~v)\neq(0,~0)}\frac{\widetilde{g}_{r,v}\left(\mathbf{P}\right)}{\alpha^{(l,t)}_{r,v}}\\\label{eq_SPCA_UB}
\overset{\bigtriangleup}{=} &\widetilde{\mathrm{UB}}^{(t)}_{\mathrm{FAP}}\left(\mathbf{P}\right),
\end{align}\label{eq_SPCA_UB1}
\end{subequations}
{with}
\begin{subequations}
\begin{align}\label{eq_alpha_lt_nv}
\alpha^{(l,t)}_{r,v}=&2\sqrt{\widetilde{g}_{r,v}\left(\mathbf{P}^{(t)}\right)+\widetilde{g}_2\left(\mathbf{U}^{(l)}\right)},\\\label{eq_alpha_lt_nv2}
\alpha^{(l,t)}=&\sum_{(r,~v)\neq(0,~0)}\frac{\widetilde{g}_{r,v}\left(\mathbf{P}^{(t)}\right)+2\widetilde{g}_2\left(\mathbf{U}^{(l)}\right)}{\alpha^{(l,t)}_{r,v}}.
\end{align}
\end{subequations}

Hence, combing \eqref{eq_SCP_P_SCA} and  \eqref{eq_SPCA_UB1}, the sub-problem to solve problem \eqref{eq_optimization_P3}  becomes
\begin{mini!}
        {\mathbf{P}}{-\mathrm{Tr}\left({\mathbf{G}^{(t)}}^T\mathbf{P}\right)+\frac{\rho}{2}\|\mathbf{P}-\mathbf{U}^{(l)}-\text{diag}\left({\boldsymbol{\sigma}}^{(l)}\right)+\mathbf{V}^{(l)}\|^2,}{\label{eq_optimization_P5}}\nonumber
        \addConstraint{ \mathrm{Tr}\left(\mathbf{P}\right) \leq P_{\max}}\label{eq_optimization_P5_2}
        \addConstraint{\widetilde{\mathrm{UB}}^{(t)}_{\mathrm{FAP}}(\mathbf{P}) \leq S_{\max},}\label{eq_optimization_P5_3}       
\end{mini!}
which is a  convex Quadratic Constrained Quadratic Programming (QCQP) problem, whose  Lagrangian  is given by
\begin{align}
\nonumber L(\mathbf{P},~u_1,~u_2)
=&-\sum_{n}\mathrm{Tr}\left({\mathbf{G}^{(t)}}^T\mathbf{P}\right)+\frac{\rho}{2}\|\mathbf{P}-\mathbf{U}^{(l)}-\text{diag}\left({\boldsymbol{\sigma}}^{(l)}\right)+\mathbf{V}^{(l)}\|^2\\\label{eq_KKT_P}
&+u_1\left( \mathrm{Tr}\left(\mathbf{P}\right) - P_{\max}\right)+u_2\left( \widetilde{\mathrm{UB}}^{(t)}_{\mathrm{FAP}}(\mathbf{P}) - S_{\max}\right),
\end{align}
where $u_1$ and $u_2$ are the non-negative Lagrange multipliers.

Taking the first-order derivative and combining \eqref{eq_SPCA_UB}, we have 
\begin{align}\nonumber
&\frac{\partial L(\mathbf{P},~u_1,~u_2)}{\partial \mathbf{P}}\\\nonumber\nonumber
=&-\mathbf{G}^{(t)}+\rho\left( \mathbf{P}-\mathbf{U}^{(l)}-\text{diag}\left(\boldsymbol{\sigma}^{(l)}\right)+\mathbf{V}^{(l)} \right) \\&+ \left[{u_1}-u_2 M(K_GM-1) \right]\mathbf{I}_{2K}\\ \label{eq_KKT_P_first_order}
&+u_2\text{diag}\left\lbrace\sum_{\footnotesize{(r,~v)\neq(0,~0)}}\left[\frac{2{\mathbf{F}}^{\mathrm{R}}_{r,v}+2{\mathbf{F}}^{\mathrm{I}}_{r,v}+4M\mathbf{I}_{2K}}{\alpha^{(l,t)}_{r,v}} \right]\text{diag}\left(\mathbf{P}\right)\right\rbrace,
\end{align}
where ${\mathbf{F}}^{\mathrm{R}}_{r,v}={\mathbf{f}^{\mathrm{R}}_{r,v}}^T {\mathbf{f}^{\mathrm{R}}_{r,v}}$ and ${\mathbf{F}}^{\mathrm{I}}_{r,v}={\mathbf{f}^{\mathrm{I}}_{r,v}}^T {\mathbf{f}^{\mathrm{I}}_{r,v}}$.

Equating  \eqref{eq_KKT_P_first_order}  to $\mathbf{0}$, we observe that, for the non-diagonal elements of $\mathbf{P}$, the stationary point is irrelevant with the multipliers $u_1$ and $u_2$, and can be directly calculated as 
\begin{align}
\label{eq_KKT_P_2}
{P}_{k,j}=&\frac{1}{\rho} {G}^{(t)}_{k,j}+{U}^{(l)}_{k,j}-{V}^{(l)}_{k,j},~~(j\neq k).
\end{align}

For the diagonal elements of  $\mathbf{P}$,  \eqref{eq_KKT_P_first_order} can be re-organized into \eqref{eq_p^D} where $\mathbf{p}=\mathrm{diag}\left(\mathbf{P}\right)$, $\mathbf{g}^{(t)}=\mathrm{diag}\left(\mathbf{G}^{(t)}\right)$, $\boldsymbol{\mu}^{(t)}=\mathrm{diag}\left(\mathbf{U}^{(t)}\right)$ and $\mathbf{v}^{(t)}=\mathrm{diag}\left(\mathbf{V}^{(t)}\right)$. After being setting into $\mathbf{0}$, \eqref{eq_p^D} gives the result  as  \begin{align}
\label{eq_KKT_P_1}
{P}_{k,k}=p_k=\left(\begin{bmatrix}
\mathbf{Q}^{(t)}\left(u_2\right)& \mathbf{1} \\
\mathbf{1} &0
\end{bmatrix}^{-1}\begin{bmatrix}
-\mathbf{q}^{(t)}\left(u_2\right)\\
 P_{\max}
\end{bmatrix}\right)_{k},
\end{align}
where $k=0,~\cdots,~2K-1$. $u_2$ is obtained by bi-section search. $\mathbf{Q}^{(t)}\left(u_2\right)$ and $\mathbf{q}^{(t)}\left(u_2\right)$ are shown in \eqref{eq_p^D}.

\begin{figure*}
\begin{align}
\label{eq_p^D}
\scriptstyle{\frac{\partial L}{\partial \mathbf{p}}
=}&\underbrace{\scriptstyle{\left\lbrace \rho\mathbf{I}_{2K}+u_2\sum_{r,v} \frac{2}{\alpha^{(l,t)}_{r,v}}\left[{\mathbf{F}}^{\mathrm{R}}_{r,v}+ {\mathbf{F}}^{\mathrm{I}}_{r,v}+2M\mathbf{I}_{2K}\right]\right\rbrace}}_{\scriptstyle{\mathbf{Q}^{(t)}\left(u_2\right)}}{\mathbf{p}}{+ {u_1}  }\underbrace{\scriptstyle{-u_2 M(K_GM-1)-\mathbf{g}^{(t)}+{\rho}\left( -\boldsymbol{\mu}^{(l)}-{\boldsymbol{\sigma}}^{(l)}+\mathbf{v}^{(l)} \right) }}_{\scriptstyle{\mathbf{q}^{(t)}\left(u_2\right)}}.
\end{align}
\noindent\makebox[\linewidth]{\rule{\paperwidth}{0.4pt}}
\vspace*{-0.4cm}
\end{figure*}

The algorithm to solve sub-problem \eqref{eq_ADMM_P} is summarized in algorithm \ref{SCP-P}. \textcolor{blue}{For sub-problem \eqref{eq_ADMM_P}, the overall SCA iterative computational complexity is $\mathcal{O}\left(K^4 \right)$. The iterative computational complexity for each step of the SCA is explicit within algorithm \ref{SCP-P}.}
\begin{algorithm}[t]
 $\textbf{Input}$: $t\leftarrow0,~\mathbf{P}^{(l+1,~0)},~ \epsilon_0,~u_{2_\mathrm{L}},~u_{2_\mathrm{U}}$\\
 $\textbf{Output}$: $\mathbf{P}^{(l+1)}$\\
  $\textbf{Repeat}$: \\
 \begin{enumerate}
    \item Compute $\mathbf{G}^{(t)} $ from   \eqref{eq_P_SCP_coeff} \textcolor{blue}{with computational complexity $\mathcal{O}\left(K^4 \right)$}, $\alpha^{(l,t)}_{r,v}$  and $\alpha^{(l,t)}$ from  \eqref{eq_alpha_lt_nv} at the operating point $\mathbf{P}^{(l+1,~t)}$ \textcolor{blue}{with computational complexity $\mathcal{O}\left(K^2 \right)$}\\
    \item Find $u_2$ by bi-section search \textcolor{blue}{with computational complexity $\mathcal{O}\left(K^2 \right)$}:\\
    \begin{enumerate}
        \item  $u_2\leftarrow 0$, $\mathbf{P}\left( 0\right)$ from  \eqref{eq_KKT_P_2} and  \eqref{eq_KKT_P_1}   \\
        \item  \textbf{If } $\widetilde{\mathrm{UB}}^{(t)}_{\mathrm{FAP}}\left(\mathbf{P}\left(u_2\right)\right)>S_{\max}$ \\
        \begin{enumerate} 
             \item[] $\mathbf{P}^{(l+1,~t+1)}\leftarrow\mathbf{P}\left(u_2\right)$ \\
         \end{enumerate} 
        \item[]\textbf{else } \\
        \begin{enumerate} 
            \item[] $\mathbf{P}^{(l+1,~t+1)}\leftarrow\mathbf{P}\left(u_2\right)$\\ 
            \item[] \textbf{while }$|\widetilde{\mathrm{UB}}^{(t)}_{\mathrm{FAP}}\left(\mathbf{P}\left(u_2\right)\right)-S_{\max}|>\epsilon_0$\\
            \begin{enumerate}
                \item[]  $u_2=(u_{2_\mathrm{L}}+u_{2_\mathrm{U}})/2$ \\
                \item[]  calculate $\mathbf{P}\left( u_2\right)$ from  \eqref{eq_KKT_P_2} and   \eqref{eq_KKT_P_1} \\
                \item[] \textbf{If } $\widetilde{\mathrm{UB}}^{(t)}_{\mathrm{FAP}}\left(\mathbf{P}\left(u_2\right)\right)<S_{\max}$  \\
               
                     \item[]  ~~$u_{2_\mathrm{U}}=u_2$\\ 
               
                \item[] \textbf{elseif} $\widetilde{\mathrm{UB}}^{(t)}_{\mathrm{FAP}}\left(\mathbf{P}\left(u_2\right)\right)>S_{\max}$  \\
                 
                     \item[] ~~$u_{2_\mathrm{L}}=u_2$\\
               
                \item[]  \textbf{end} \\
            \end{enumerate}   
            \item[]  \textbf{end}\\         
        \end{enumerate}  
         \item[] \textbf{end} \\
    \end{enumerate}
    \item Compute $\mathbf{P}^{(l+1,~t+1)}$ combining   \eqref{eq_KKT_P_2} and  \eqref{eq_KKT_P_1} \textcolor{blue}{with computational complexity $\mathcal{O}\left(K^3 \right)$}\\
    \item $t \leftarrow t+1$; $\mathbf{P}^{(l+1)} \leftarrow \mathbf{P}^{(l+1,~t+1)}$\\
    \item Quit if $\|\mathbf{P}^{(l+1,~t+1)}-\mathbf{P}^{(l+1,~t)}\|< \epsilon_0\|\mathbf{P}^{(l+1,~t+1)}\|$
 \end{enumerate}
\caption{SCA for $\mathbf{P}^{(l+1)}$}
\label{SCP-P}
\end{algorithm}

\subsubsection{Updating $\mathbf{U}^{(l+1)}$ and ${\boldsymbol{\sigma}}^{(l+1)}$}
Similarly, the sub-problem to solve    \eqref{eq_ADMM_U} is 
\begin{mini!}
        {\mathbf{U},~\boldsymbol{\sigma}}{-z_{\mathrm{DC}}\left(\mathbf{P}^{(l+1)},{\mathbf{U}}\right)+\frac{\rho}{2}\|-\mathbf{P}^{(l+1)}+{\mathbf{U}}+\text{diag}\left(\boldsymbol{\sigma}\right)-{\mathbf{V}}^{(l)}\|^2,}{\label{eq_optimization_P6}}{\label{eq_optimization_P6_1}}
        \addConstraint{\mathbf{U} \geq \mathbf{0}}\label{eq_optimization_P6_2}
        \addConstraint{\boldsymbol{\sigma}\succeq \mathbf{0}}\label{eq_optimization_P6_3}
        \addConstraint{R(\boldsymbol{\sigma}) \geq C_{\min}}\label{eq_optimization_P6_4}
         \addConstraint{\widetilde{\mathrm{UB}}_{\mathrm{FAP}}(\mathbf{P}^{(l+1)},~\mathbf{U}) \leq S_{\max}}\label{eq_optimization_P6_5},
\end{mini!}
whose concave part in the objective function in  \eqref{eq_optimization_P6_1} is still substituted by a linear approximation by using SCA. Specifically, at the $t^{\mathrm{th}}$ SCA iteration, $-z_{\mathrm{DC}}\left(\mathbf{P}^{(l+1)},{\mathbf{U}}\right)$ in   \eqref{eq_optimization_P6_1} becomes 
\begin{align}
\nonumber\label{eq_SCP_P}{z_{\mathrm{DC}}}^{(t)}\left( \mathbf{U}\right)
=&{z_{\mathrm{DC},~0}}^{(t)}+\mathrm{Tr}\left({\mathbf{J}^{(t)}}^T\mathbf{U}\right)-\sum_{n=K_G}^{K'-1}2\mathrm{Tr}\left( \mathbf{A}_n\mathbf{U}\mathbf{A}_n\mathbf{U}\right)\\
&-
\sum_{n=0}^{K_G-1}2\left[\mathrm{Tr}\left( \mathbf{E}_n\mathbf{U}\mathbf{E}_n\mathbf{U}\right)+\mathrm{Tr}\left( \widetilde{\mathbf{E}}_n\mathbf{U} \widetilde{\mathbf{E}}_n\mathbf{U}\right)\right],
 \end{align}
 where $ \mathbf{J}^{(t)}$ is given in  \eqref{eq_taylor_coeff_u}. ${z_{\mathrm{DC},~0}}^{(t)}$ is the Taylor constant to be omitted. Note that $\mathbf{A}_n$ is positive semi-definite from the definition and $\mathbf{E}_n$/$\widetilde{\mathbf{E}}_n$ are positive semi-definite as proved in Appendix \ref{appen2}.
 \begin{figure*}
 \begin{align}
 \nonumber
\scriptstyle{\mathbf{J}^{(t)}=}&\scriptstyle{\sum_{k=0}^{K_G-1}2k_2\mathbf{D}_n^T+ {3k_4} \left[\mathrm{Tr}\left(\mathbf{B}_{n}\mathbf{P}^{(l+1)}+2\mathbf{D}_{n}\mathbf{U}^{(t)} \right)\mathbf{D}_{n}^T+\mathrm{Tr}\left(\widetilde{\mathbf{B}}_{n}\mathbf{P}^{(l+1)}+2\widetilde{\mathbf{D}}_{n}\mathbf{U}^{(t)} \right)\mathbf{D}_{n}^T+3\sigma_{\mathrm{n_P}}^2\mathbf{D}_{n}^T+3\sigma_{\mathrm{n_P}}^2\widetilde{\mathbf{D}}_n^T\right]}\\\nonumber
 &\scriptstyle{+\sum_{k=0}^{K_G-1} {3k_4}\left[2\mathbf{D}^T_{n}\mathbf{U}^{(t)}\mathbf{D}^T_{n}+\mathbf{B}_{1,~n}\mathbf{U}^{(t)}\mathbf{B}_{2,~n}+\mathbf{B}_{2,~n}\mathbf{U}^{(t)}\mathbf{B}_{1,~n}+\mathbf{D}_{n}^T\mathbf{P}^{(l+1)}\mathbf{B}_{1,~n}+\mathbf{D}_{n}\mathbf{P}^{(l+1)}\mathbf{B}_{2,~n}+\mathbf{B}_{1,~n}\mathbf{P}^{(l+1)}\mathbf{D}_{n}+\mathbf{B}_{2,~n}\mathbf{P}^{(l+1)}\mathbf{D}_{n}^T\right]}\\\label{eq_taylor_coeff_u}
&\scriptstyle{+\sum_{k=0}^{K_G-1} {3k_4}\left[2\widetilde{\mathbf{D}}^T_{n}\mathbf{U}^{(t)}\widetilde{\mathbf{D}}^T_{n}+\widetilde{\mathbf{B}}_{1,~n}\mathbf{U}^{(t)}\widetilde{\mathbf{B}}_{2,~n}+\widetilde{\mathbf{B}}_{2,~n}\mathbf{U}^{(t)}\widetilde{\mathbf{B}}_{1,~n}+\widetilde{\mathbf{D}}_{n}^T\mathbf{P}^{(l+1)}\widetilde{\mathbf{B}}_{1,~n}+\widetilde{\mathbf{D}}_{n}\mathbf{P}^{(l+1)}\widetilde{\mathbf{B}}_{2,~n}+\widetilde{\mathbf{B}}_{1,~n}\mathbf{P}^{(l+1)}\widetilde{\mathbf{D}}_{n}+\widetilde{\mathbf{B}}_{2,~n}\mathbf{P}^{(l+1)}\widetilde{\mathbf{D}}_{n}^T\right]}.
 \end{align}
 \noindent\makebox[\linewidth]{\rule{\paperwidth}{0.4pt}}
 \vspace*{-0.4cm}
\end{figure*}

The constraint in  \eqref{eq_optimization_P6_5} is a decrease function with respect to $\|\text{diag}\left(\mathbf{U}\right)\|^2$, and can be equivalently transformed into
\begin{equation}
\label{eq_UB_U_transform}
\|\text{diag}\left(\mathbf{U}\right)\|^2 \geq r^{(l)},
\end{equation}
where $r^{(l)}$ can be found by bisection search such that ${\sum_{(r,~v)\neq (0,~0)}-\mathrm{Tr}\left(\mathbf{P}\right)+\sqrt{\widetilde{g}_{r,v}\left(\mathbf{P}\right)-2r^{(l)}}}=S_{\max}$.

 \eqref{eq_UB_U_transform} is still a concave non-linear constraint, but can be handled by introducing a parametric vector $\mathbf{v}$ and utilizing the inequality 
\begin{align}
\label{eq_UB_U_radar}
\|\text{diag}\left(\mathbf{U}\right)\|^2\geq \frac{\left(\mathbf{v}^T \|\text{diag}\left(\mathbf{U}\right)\|\right)^2}{\|\mathbf{v}\|^2},
\end{align}
which holds if and only if $\text{diag}\left\lbrace \mathbf{U}\right\rbrace_k/v_k = c$ ($c$ constant) for $\forall\:\: k$. Hence, the constraint in   \eqref{eq_UB_U_transform} can be substituted by   \eqref{eq_UB_U_radar}, taking $\mathbf{v}$  as the diagonal elements of the optimal $\mathbf{U}$ from the previous round in each SCA iteration. The algorithm that introduces auxiliary variables to form an achievable convex upper bound is named parametric SCA (PSCA) \cite{beck2010sequential}.

Combining with   \eqref{eq_UB_U_radar}, at the $t^{\mathrm{th}}$ iteration, we have
\begin{mini!}
        {\mathbf{U},~\boldsymbol{\sigma}}{{z_{\mathrm{DC}}}^{(t)}\left(\mathbf{U}\right)+\frac{\rho}{2}\|-\mathbf{P}^{(l+1)}+{\mathbf{U}}+\text{diag}\left(\boldsymbol{\sigma}\right)-{\mathbf{V}}^{(l)}\|^2}{\label{eq_optimization_P8}} {\label{eq_optimization_P8_1}}
         \addConstraint{\frac{{\mathbf{v}^{(t)}}^T \text{diag}\left(\mathbf{U}\right)}{\sqrt{\|\mathbf{v}^{(t)}\|^2}} \geq \sqrt{r^{(l)}}}\label{eq_optimization_P8_5}
         \addConstraint{ \eqref{eq_optimization_P6_2}, ~\eqref{eq_optimization_P6_3}, ~\text{and}~
 \eqref{eq_optimization_P6_4},}\label{eq_optimization_P8_3}
\end{mini!}
where $\mathbf{v}^{(t)}=\text{diag}\left(\mathbf{U}^{(t)}\right)$ and which is a convex programming that can be solved by cvx.

The algorithm to solve sub-problem \eqref{eq_ADMM_U} is summarized in algorithm \ref{SCA-PSCA}. \textcolor{blue}{Similarly, for sub-problem \eqref{eq_ADMM_U}, the overall SCA iterative computational complexity is $\mathcal{O}\left(K^4 \right)$. The iterative computational complexity for each step of the SCA is explicit within algorithm \ref{SCA-PSCA}.} The whole algorithm is summarized in Algorithm \ref{ADMM}.

\begin{algorithm}[t]
  $\textbf{Input}$: $t\leftarrow0,~\left(\mathbf{U}^{(l,~0)},{\boldsymbol{\sigma}}^{(l,~0)}\right),~ \epsilon_0$;\\
 $\textbf{Output}$: $\left(\mathbf{U}^{(l+1)}, ~{\boldsymbol{\sigma}}^{(l+1)}\right)$;\\
  $\textbf{Repeat}$: \\
  \begin{enumerate}
\item Compute $\mathbf{J}^{(t)} $ from  \eqref{eq_taylor_coeff_u} at $\left(\mathbf{U}^{(l,t)},{\boldsymbol{\sigma}}^{(l,t)}\right)$ \textcolor{blue}{with computational complexity $\mathcal{O}\left(K^4 \right)$};\\
\item Compute $\mathbf{v}^{(t)}=\text{diag}\left(\mathbf{U}^{(t)}\right)$\\
\item Compute $\left(\mathbf{U}^{(l,t)},~{\boldsymbol{\sigma}}^{(l,t)}\right)$ from \eqref{eq_optimization_P8} using cvx \textcolor{blue}{with computational complexity $\mathcal{O}\left(K^3 \right)$ (assume interior-point method \cite[Chapter 1.3]{boyd2004convex})}\\
\item $t\leftarrow t+1$; $\left(\mathbf{U}^{(l+1)},~{\boldsymbol{\sigma}}^{(l+1)}\right) \leftarrow \left(\mathbf{U}^{(l+1,~t+1)},~{\boldsymbol{\sigma}}^{(l+1,~t+1)}\right)$\\
\item Quit if 
$\|\mathbf{U}^{(l+1,~t+1)}-\mathbf{U}^{(l+1,~t)}\|< \epsilon_0\|\mathbf{U}^{(l+1,~t+1)}\|$
\end{enumerate}
\caption{SCA-PSCA for $\left(\mathbf{U}^{(l+1)},~{\boldsymbol{\sigma}}^{(l+1)}\right)$}
\label{SCA-PSCA}
\end{algorithm}

\begin{algorithm}[t]
 $\textbf{Input}$: $l\leftarrow 0,~k\leftarrow 0,~{\mathbf{P}}^{(0)},~{\mathbf{U}}^{(0)},~{\boldsymbol{\sigma}}^{(0)},~\epsilon_S>0,~\rho>0$\\
 $\textbf{Output}$: ${\mathbf{p}}^{\star},~{\boldsymbol{\mu}}^{\star},~{\boldsymbol{\sigma}}^{\star}$\\
 \begin{enumerate}
     \item $\textbf{While } k<20$ \\
     \begin{enumerate}
         \item[] $\textbf{While } \|\mathbf{P}^{(l+1)}-\mathbf{P}^{(l+1)}\|< \epsilon_0\|\mathbf{P}^{(l+1)}\|$ \\
         \begin{enumerate}
            \item[] Update $\mathbf{P}^{(l+1)}$ by Algorithm \ref{SCP-P}\\
            \item[] Update $\left(\mathbf{U}^{(l+1)},~{\boldsymbol{\sigma}}^{(l+1)}\right)$ by Algorithm \ref{SCA-PSCA}\\
            \item[] Update $\mathbf{V}^{(l+1)}$ in  \eqref{eq_ADMM_v} \\
            \item[] $l\leftarrow l+1$\\
         \end{enumerate}    
         \item[] $\textbf{end}$\\
     \end{enumerate}
     \item[] $\textbf{end}$
     \item Obtain $\boldsymbol{\mu}^{\star}$ from   ${\mathbf{U}}^{\star}$ using Gaussian randomization\cite{luo2010semidefinite}\\
     \item ${p}^{\star}_k\leftarrow {{\mu}^{\star}_k}^2+ {{\sigma}}^{\star}_k$, ~ $k=(0,~\cdots,~2K-1)$
 \end{enumerate}
\caption{ADMM}
\label{ADMM}
\end{algorithm}

\section{Simulation Results}
\label{sec_sim}

In this section, we simulate the performance of the ISCAP system and evaluate its C-P region (maximal harvested power as a function of the achievable rate constraint, for a fixed aISPLD\footnote{The aISPLD constraint is normalized when plotting the simulations.} constraint) and its S-P region (maximal harvested power as a function of the aISPLD constraint, for a fixed achievable rate constraint).

Throughout the simulations, the performance of the ISCAP system is evaluated under a Wi-Fi like scenario with $f_0=5.18$ GHz. At the transmitter, assume an OFDM signal with bandwidth $B=30$ MHz, $K=8$ subcarriers, $K_G=4$ CP pulses and $40$ dBm transmit power. The communication channel is an NLOS model for indoor WiFi scenario adopted from Model B in \cite{channel}\footnote{\textcolor{blue}{The channel model is widely adopted and recognized in literature \cite{clerckx2016waveform, varasteh2019swipt}, and shares significant similarities with the state-of-art 3GPP wideband (indoor office/industry) channel model in \cite[Tapped Delay Line models]{channel2}, both providing data on multi-paths delay and degradation while incorporating Rayleigh distribution for the small-scale fading.}},  with a $108$ dB path-loss (around 1 km), $-110$ dBW noise power, and a $0$ dB communication SNR over simulations. The powering channel is similarly simulated as the communication channel but with a $58$ dB path-loss (around 10 m), with a $-108$ dBW noise power. For the sensing function, we assume an average $-20$  dB radar receiver SNR at over simulations.

\subsection{S-P region}
\label{subsec:sp_region}
\begin{figure*}[t]
\centering
\includegraphics[scale=0.6]{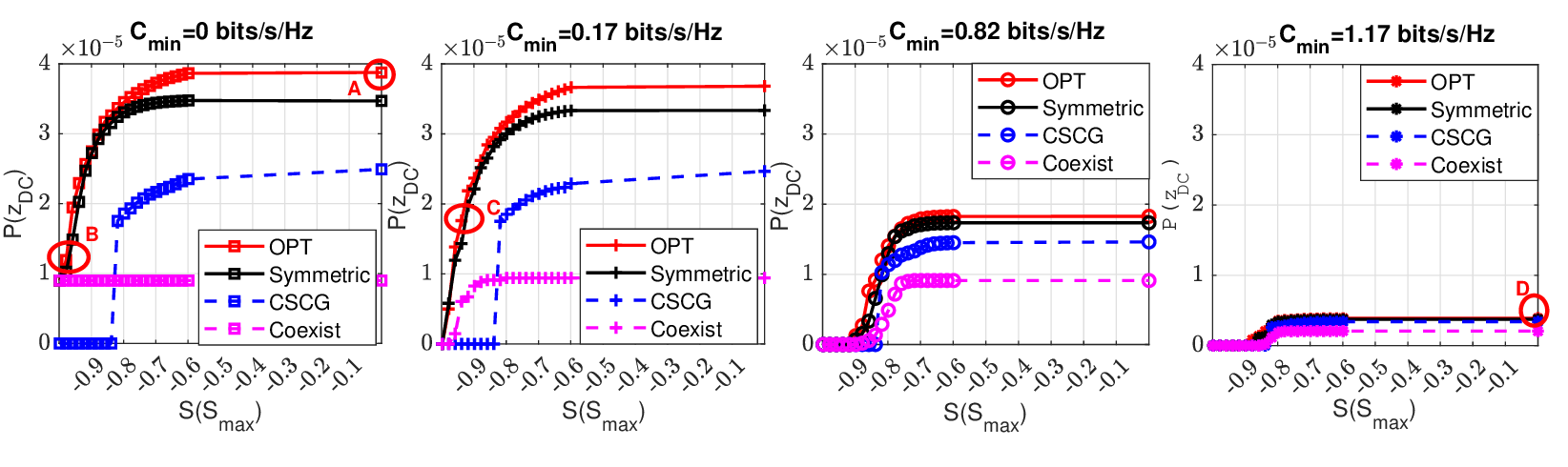}
\caption{The average S-P region given different achievable rate constraints ($C_{\min}=0,~0.12,~0.47,~0.82,~1.17$ bits/s/Hz from left to right). `OPT' (red) refers to the optimized asymmetric complex Gaussian input distribution from Section \ref{section_opt}. `Symmetric' (black) refers to the optimized symmetric complex Gaussian input distribution. `CSCG' (blue) refers to the optimized CSCG input distribution. `Coexist' (magenta) refers to the non-co-design signal superposed by the optimal communication and sensing signals through power-splitting. Generally, 'OPT' achieves the largest S-P region, followed by the co-designed 'Symmetric' and 'CSCG', which all outperform the 'Coexist' signal. $P(Z_{\mathrm{DC}})=0$ means that the exerted rate and aISPLD constraints are not feasible  at the point.}
\label{fig_R_E}
\end{figure*}

\begin{figure*}[!t]
\centering
\subfloat[]{\includegraphics[width=3.0in]{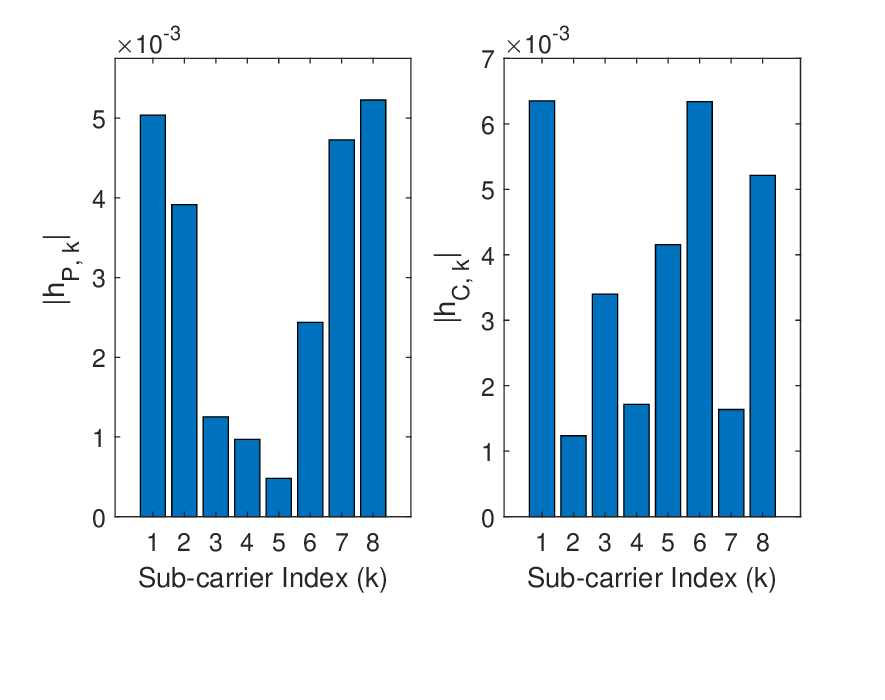}%
\label{fig_R_E_Channel}}
\hfil
\subfloat[]{\includegraphics[width=3.0in]{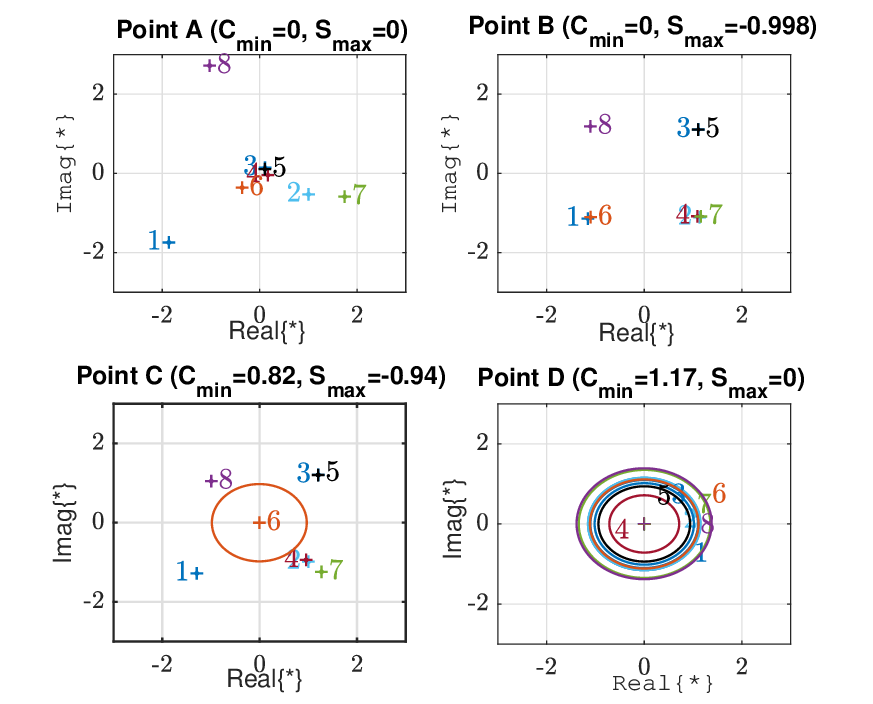}%
\label{fig_R_E_points}}
\hfil
\caption{(a) Left: Magnitude response of one powering channel realization, Right: Magnitude response of one communications channel realization; (b) For the channel realizations in (a), the 'OPT' input distribution corresponding to points A, B, C, and D in Fig. \ref{fig_R_E}. For each subcarrier (numbered 1 through 8), its complex Gaussian distribution is represented by an ellipse whose center is the mean, and whose width and height correspond to the variance of the real and imaginary parts, respectively. For detailed insights on the figures on this page, see Section~\ref{subsec:sp_region}.}
\label{fig_R_E_one}
\end{figure*}

Fig. \ref{fig_R_E} plots the average S-P region over 300 channel realizations, given different constraints on the communication achievable rate. We compare three co-design input distributions; namely, the asymmetric complex Gaussian from Section \ref{section_opt} (red curve with legend `OPT'), the symmetric complex Gaussian (black curve with legend `Symmetric'),  the CSCG (blue curve with legend 'CSCG')\footnote{The optimal 'Symmetric' input and the optimal 'CSCG' input are also achieved from the optimization problem in section \ref{section_opt}. For the optimal 'Symmetric' input, we substitute $\mu^{\mathrm{I}}_{k}=\mu^{\mathrm{R}}_{k}$, $\sigma^{\mathrm{I}}_{k}=\sigma^{\mathrm{R}}_{k}$ in problem \eqref{eq_optimization_P1} and only needs to optimize $2K$ variables rather than $4K$. Similarly, for the optimal 'CSCG' input, we only need to optimize $\sigma^{\mathrm{I}}_{k}=\sigma^{\mathrm{R}}_{k}$ with $\mu^{\mathrm{I}}_{k}=\mu^{\mathrm{R}}_{k}=0$.}, and the coexisting power-splitting input distribution (magenta curve with legend 'Coexist'\footnote{The 'Coexist' input satisfies the same transmit power constraint as the three co-design inputs.  In particular, the variance of the 'Coexist' input is decided by the communication function (the variance that satisfies the achievable rate with minimal power) and the mean of the 'Coexist' input is the optimal sensing input using the remaining power. \textcolor{blue}{Consequently, the 'Coexist' waveform is a beneficial waveform for the ISAC scenario.}}).  Figs.~\ref{fig_R_E_Channel} and \ref{fig_R_E_points} are companion figures to Fig. \ref{fig_R_E}. Specifically, Fig. \ref{fig_R_E_points} plots the 'OPT' input distributions corresponding to points 'A', 'B', 'C' and 'D' in Fig.~\ref{fig_R_E} for channel realizations given by Fig. \ref{fig_R_E_Channel}. The insights from this collection of figures are as follows:
\begin{itemize}
    \item Point 'A' in Fig.~\ref{fig_R_E} refers to the scenario without achievable rate and aISPLD constraints, i.e., a pure powering system. From the corresponding plot in Fig. \ref{fig_R_E_points}, we observe that all the power is allocated to the symbol means of the subcarrier, proportional to the strength of the powering channel in Fig. \ref{fig_R_E_Channel}. For instance, subcarriers 8 and 1 have the two strongest powering channels (large values of $|h_{P,k}|$) and hence, the largest amount of power allocated to them (i.e., farthest from the origin). Furthermore, the optimal symbol means of the real and imaginary parts are unequal, in line with the asymmetric complex Gaussian distribution,  {highlighting the powering system's performance dependence on having different real and imaginary parts of the symbol means that are adaptive to the powering channels, as it has mentioned in Footnote \ref{footnote_symmetric} and Remark \ref{remark_WPT}.} Finally, we see from Fig.~\ref{fig_R_E} that for fixed achievable rate and aISPLD constraints, the asymmetric complex Gaussian input distribution leads to higher harvested power than the other three symmetric input distributions.

    \item From 'A' to 'B', a stricter aISPLD constraint is imposed (from $S_{\max}=0$ to $S_{\max}=-0.998$, with the lowest achievable aISPLD being $S_{\max}=-1$). However, like 'A', there is no achievable rate constraint. Hence, 'B' corresponds to an input distribution designed for \emph{integrated sensing and power transfer}. The resulting 'OPT' input distribution still allocates all of the available transmit power to the symbol means, but with a more uniform power allocation across subcarriers. This is because the lowest value of the aISPLD metric in \eqref{eq_aISPLD} is achieved when the OFDM symbols have all of the power allocated to the symbol means and have the same magnitude of the symbol means across sub-carriers \cite{2022ISAC}.   
    {Indeed, Fig. \ref{fig_R_E} shows that the lowest value of aISPLD can be achieved by both asymmetric ('OPT') and symmetric ('Symmetric') inputs -- the non-zero power $P(z_{\mathrm{DC}})$ starts from the same value of $S(S_{\mathrm{max}})$ axis in Fig. \ref{fig_R_E} (they share the same feasible region of the sensing and communication constraints). The takeaway is that, while the (complex) symbol mean is essential for sensing, the sensing's performance is not affected by either having identical or different means on the real and imaginary symbol parts, as long as the total power allocated to the symbol means is fixed, which is in sharp contrast with the powering system's preference as it has been mentioned in Remark \ref{remark_sensing}. }

    \item From 'B' to 'C', the aISPLD is looser (from $-0.998$ to $=-0.94$), but the achievable rate constraint, $C_{\min}$, is tighter (from $0$ to $0.82$ bits/s/Hz)\footnote{The highest achievable rate for the channel in Fig. \ref{fig_R_E_Channel} is $1.23$ bits/s/Hz.}. Since the sensing and communications constraints are non-trivial, 'C' corresponds to an input distribution designed for ISCAP. Compared with 'B', the 'OPT' input distribution of 'C' still shows a tendency for uniform power allocation across subcarriers to satisfy the aISPLD constraint, except that at subcarrier $6$ (strongest communication channel in Fig. \ref{fig_R_E_Channel}) is the allocated power is shifted from the mean to the variance in order to satisfy the achievable rate constraint. 

    \item From 'C' to 'D', the aISPLD is further loosened (to $S_{\max}=0$ the same as 'A'), whereas the achievable rate constraint is further tightened (to $1.17$ bits/s/Hz). \textcolor{blue}{Thus, point 'D' corresponds to an input distribution designed for WIPT, whose harvested power exceeds all the other points on the 'OPT' curve since no sensing constraint is exerted there.} Due to the large value of $C_{\rm min}$, the available power is allocated solely to the symbol variance of each subcarrier, largely in proportion to the strength of the communication channel. For instance, subcarriers 1 and 6 have the two strongest communication channels (large values of $|h_{C,k}|$) and hence, the largest amount of power allocated to them (i.e., larger ellipses). Once the achievable rate constraint is satisfied, the remaining power is distributed to subcarriers with strong powering channels (e.g., subcarrier $7$, which also has a large ellipse). Subcarriers with weak communications and powering channels (e.g., subcarrier 4) have no power allocated to them.
\end{itemize} 

 
To summarize, {Fig. \ref{fig_R_E} demonstrates that our proposed non-zero mean asymmetric 'OPT' distribution achieves the largest S-P region, followed by the 'Symmetric' and the 'CSCG' input distributions. In particular, due to its asymmetric input, the gap between the 'OPT' and other S-P regions enlarges as the achievable rate and/or aISPLD constraint is loosened. Generally, the three co-design inputs (especially the 'OPT' and the 'Symmetric') outperform the 'Coexist'  for any achievable rate constraint.} The comparison between the different points in Fig. \ref{fig_R_E} indicates the following two general trade-offs between the three functions in the ISCAP system:
\begin{itemize}
    \item[1)] the trade-off between power allocation to the symbols mean and variance -- the former is beneficial for sensing and powering (especially with frequency-flat channels) while the latter is essential for communications; and
    \item[2)] for power allocation to the symbol mean in particular, powering prefers an allocation that is asymmetric and proportional to the strength of powering channel, while sensing favors uniform and symmetric power allocation.
\end{itemize} 

\subsection{C-P region}
\begin{figure*}[ht]
\centering
\subfloat[]{\includegraphics[width=3.0in]{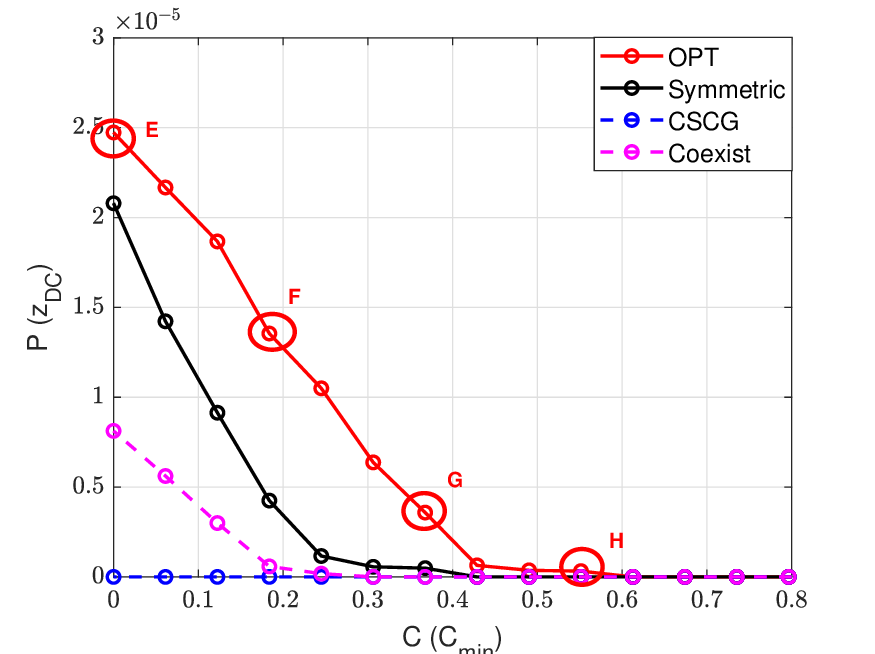}%
\label{fig_C_E}}
 \subfloat[]{\includegraphics[width=3.0in]{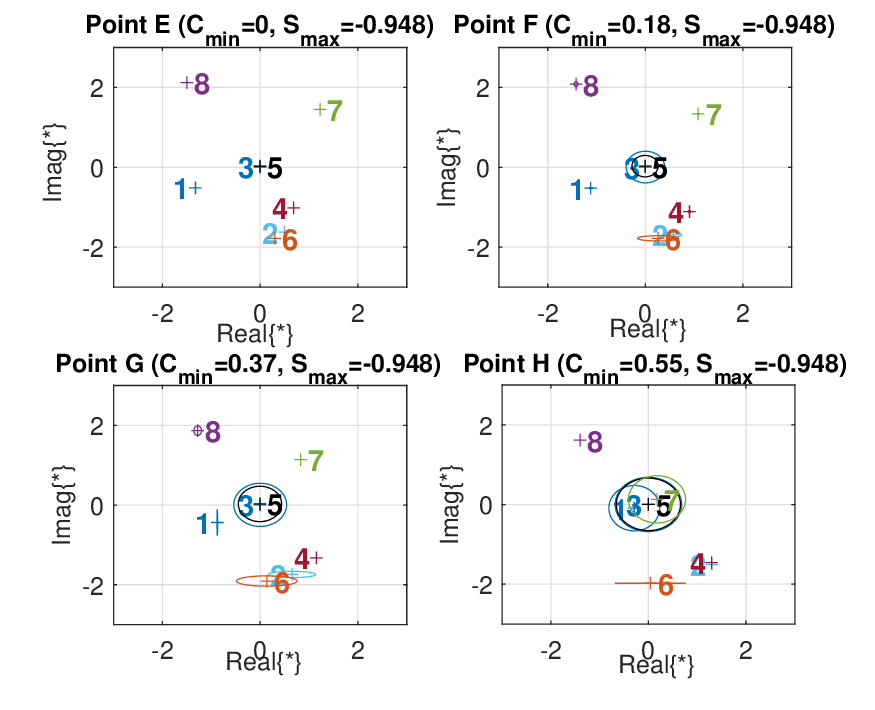}%
\label{fig_C_E_points}}
 \caption{(a) The average C-P region given $S_{\max}=-0.948$,  comparing the 'OPT', 'Symmetric', 'CSCG' and 'Coexist' inputs. (b) The 'OPT' input constellation of points E, F, G and H in Fig. \ref{fig_C_E} corresponds to the channel realizations in Fig. \ref{fig_R_E_Channel}.}
\label{fig_C_E_1}
\end{figure*}

For further verification, we plot the average C-P region of the ISCAP system in Fig. \ref{fig_C_E} given a fixed non-trivial aISPLD constraint of $S_{\max}=-0.948$. {Similarly, four optimal input distributions on the C-P region, 'E', 'F', 'G' and 'H' are plotted in Fig. \ref{fig_C_E_points}, corresponding to the same channel realizations in Fig. \ref{fig_R_E_Channel}. The takeaways are concluded as:}

 \begin{itemize}
 \item Fig. \ref{fig_C_E} reveals a smaller performance gap between different inputs given stricter constraints on the achievable rate, since they all converge to a CSCG input (i.e., $C_{\min}\geq0.55$ bits/s/Hz) for the best communications. On the other hand, given loose achievable rate constraints, the 'OPT' input outperforms the other inputs significantly because of its flexibility to have asymmetric inputs catering to the preference of powering,  {i.e., the constellations for 'E', 'F', 'G' and 'H' have different symbol means and variance for the real and imaginary parts (e.g., subcarrier 6 with a horizontal ellipse centering around $\mu_{6}=-2j$). } In terms of the power allocation tendency, from 'E' to 'H' in Fig. \ref{fig_C_E_points}, we observe that increasing achievable rate constraints ($C_{\min}=\{0,0.18,0.37,0.55\}$ bits/s/Hz) forces more power allocated to the symbol variance, prior to the subcarriers with strong communication channels (Subcarrier 6 for example).

\item   Comparing  'E' in Fig. \ref{fig_C_E_points} with 'B'  in Fig. \ref{fig_R_E_points} (both exerted no communication constraint and corresponding to the integrated sensing and power transfer), all of their power is allocated to the symbol means. However,  'E' is less uniform than B because of looser aISPLD constraints ($S_{\max}=-0.948$ for 'E' over $S_{\max}=-0.998$ for 'B'), and is more adaptive to powering channels by allocating more power to Subcarriers 1, 7 and 8 (the strongest powering channels).  The relatively strict aISPLD of 'E' indicates that good sensing can also be achieved even with a relatively dispersive power allocation if the power is mainly allocated to the mean of the Gaussian input. On the contrary, when comparing 'E' and 'H', given the additional communication constraint at 'H',  the total power assigned to each subcarrier at 'H' has to be more uniform than 'E' to satisfy the same aISPLD constraint.

\end{itemize}

To sum up, { Fig. \ref{fig_C_E} again highlights the performance gain of the co-design inputs, especially the proposed 'OPT' input compared with the 'Coexist' input. }  In addition, with both constraints on achievable rate and aISPLD, part of the subcarriers are mainly allocated power to the symbol variance while the others are mainly allocated power to the symbol mean, forming a relatively uniform power allocation across subcarriers dependent on the strictness of the aISPLD constraint. Once satisfying the communication and sensing constraints, as much power as possible is allocated to the symbol mean of the sub-carries with strong powering channels, i.e., Subcarriers $1$ and $8$. Such an input tendency, interacted by all three functions, is hard to obtain by a simple power-splitting input, hence highlighting the significance of input signal co-design in  ISCAP.

\begin{figure}[ht]
\centering
\includegraphics[width=3.0in]{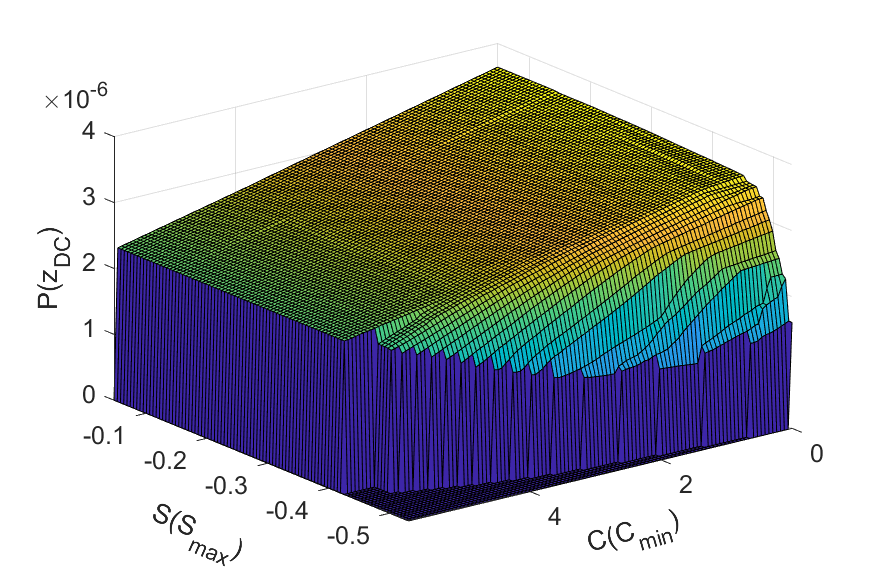}
\caption{S-C-P region of the 'OPT' input with $30$ dBm transmit power.}
\label{fig_R_C_E}
\end{figure}

The overall S-C-P region of the 'OPT' input corresponding to the channels in Fig. \ref{fig_R_E_Channel} is given in Fig. \ref{fig_R_C_E} for more details. Understandably, looser communication/sensing constraints give the flexibility to generate more energy.

\section{Conclusions}
\label{sec_con}
This paper investigates an ISCAP system, where a wideband OFDM signal is used to simultaneously serve the functions of sensing, communications, and powering. The performance of each function is evaluated given an OFDM signal with Gaussian CSs, whose input distribution across subcarriers is optimized to maximize the harvested power while satisfying the communication and sensing constraints. Simulations first verify the performance gain of the ISCAP system using the proposed input distribution over the power-splitting signals in a co-existing scenario, i.e., a larger harvested power with the same communication and sensing constraints. Further simulations and analysis reveal how the optimal input distribution trades off between the three functions: a) Firstly, powering favors asymmetric power allocation according to the powering channels, sensing allocates power uniformly to the symbol mean of each subcarrier, and communications allocates power to the symbol variance of subcarriers following a water-filling way; b) Secondly, in a general ISCAP set-up with both communication and sensing constraints, part of the subcarriers have a high symbol variance for the achievable rate while the rest subcarriers have high symbol mean, forming a relatively uniform power allocation across subcarriers to meet the aISPLD constraints. Given a looser aISPLD constraint, the input spectrum can become more powering-beneficial by allocating power asymmetrically to the symbol mean of subcarriers, adaptive to the powering channel.

\appendices
\section{Proof of   \eqref{eq_mean_OFDM} - \eqref{eq_4th_mean2_OFDM}}
\label{appen1}
For the second-order statistics in the OFDM data duration in  \eqref{eq_mean_OFDM}, we have 
\begin{subequations}
\begin{align}\nonumber
\mathbb{E}\left\lbrace |y[n,~m]|^2\right\rbrace
=&\mathbb{E}\left\lbrace\mathbf{\overline{x}}_m^T\mathbf{A}^{\mathrm{R}}_{n}\mathbf{\overline{x}}_m+2w_R[n,~m]{\mathbf{a}^{\mathrm{R}}_n}^T\mathbf{\overline{x}}_m+\mathbf{\overline{x}}_m^T\mathbf{A}^{\mathrm{I}}_{n}\mathbf{\overline{x}}_m\right\rbrace\\\nonumber
&+\mathbb{E}\left\lbrace2w_{\mathrm{I}}[n,~m]{\mathbf{a}^{\mathrm{I}}_n}^T\mathbf{\overline{x}}_m+|w[n,~m]|^2\right\rbrace\\
=&\mathbb{E}\left\lbrace\mathbf{\overline{x}}_m^T\mathbf{A}_{n}\mathbf{\overline{x}}_m\right\rbrace=\mathrm{Tr}\left(\mathbf{A}_n\mathbf{P} \right)+\sigma_{\mathrm{n_P}}^2,
\end{align}    
\end{subequations}
where $\mathbf{A}^{\mathrm{R/I}}_{n}=\mathbf{a}^{\mathrm{R/I}}_{n}{\mathbf{a}^{\mathrm{R/I}}_{n}}^T$ is positive semi-definite.

For the fourth-order statistics in   \eqref{eq_4th_mean_OFDM}, define $w_\mathrm{R/I}[n,~m]$ as the real/imaginary part of noise $w[n,~m]$ respectively, and we have 
\begin{subequations}
\begin{align}
\nonumber\mathbb{E}\left\lbrace |y[n,~m]|^4\right\rbrace
=&\mathbb{E}\left\lbrace \left(\mathbf{\overline{x}}_m^T\mathbf{A}_{n}\mathbf{\overline{x}}_m\right)^2\right\rbrace+6\sigma_{\mathrm{n_P}}^2\mathbb{E}\left\lbrace \mathbf{\overline{x}}_m^T\mathbf{A}_{n}\mathbf{\overline{x}}_m\right\rbrace+3\sigma_{\mathrm{n_P}}^4\\\nonumber\label{eq_4th_mean_OFDM_App2_2_App}
=&\mathrm{Tr}^2\left({\mathbf{A}_{n}}\mathbf{P}\right)+6\sigma_{\mathrm{n_P}}^2\mathrm{Tr}\left({\mathbf{A}_{n}}\mathbf{P}\right)+2\mathrm{Tr}\left({\mathbf{A}_{n}}\mathbf{P}{\mathbf{A}_{n}}\mathbf{P}\right)\\&-
2\mathrm{Tr}\left({\mathbf{A}_{n}}\mathbf{U}{\mathbf{A}_{n}}\mathbf{U}\right)+3\sigma_{\mathrm{n_P}}^4,
\end{align}
\end{subequations}
with $\mathbf{A}_n=\mathbf{A}^{\mathrm{R}}_{n}+\mathbf{A}^{\mathrm{I}}_{n}$ and where  \eqref{eq_4th_mean_OFDM_App2_2_App} is achieved by
\begin{subequations}
\begin{align}
\nonumber&\mathbb{E}\left\lbrace \left(\mathbf{\overline{x}}_m^T\mathbf{A}_{n}\mathbf{\overline{x}}_m\right)^2\right\rbrace\\
=&\mathbb{E}^2\left\lbrace \mathbf{\overline{x}}_m^T\mathbf{A}_{n}\mathbf{\overline{x}}_m\right\rbrace+\mathrm{Var}\left\lbrace   \mathbf{\overline{x}}_m^T\mathbf{A}_{n}\mathbf{\overline{x}}_m\right\rbrace\\
=&\mathrm{Tr}^2\left({\mathbf{A}_{n}}\mathbf{P}\right)+2\mathrm{Tr}\left({\mathbf{A}_{n}}\boldsymbol{\sigma}\mathbf{A}_{n}\boldsymbol{\sigma} \right)+4\boldsymbol{\mu}^T\mathbf{A}_{n}\boldsymbol{\sigma}\mathbf{A}_{n}\boldsymbol{\mu}\\\label{eq_ex_1_app1}
=&\mathrm{Tr}^2\left({\mathbf{A}_{n}}\mathbf{P}\right)+2\mathrm{Tr}\left({\mathbf{A}_{n}}\mathbf{P}{\mathbf{A}_{n}}\mathbf{P}\right)-2\mathrm{Tr}\left({\mathbf{A}_{n}}\mathbf{U}{\mathbf{A}_{n}}\mathbf{U}\right),
\end{align}
\end{subequations}
where $\mathrm{Var}\left\lbrace * \right\rbrace$ is the variance function.

\section{Proof of \eqref{eq_mean_CP} - \eqref{eq_4th_CP2}}
\label{appen2}
For the second-order statistics in the CP duration, we have  
\begin{subequations}
\begin{align}\nonumber
&\mathbb{E}\left\lbrace |y[n,~m]|^2\right\rbrace\\\nonumber
=&\mathbb{E}\left\lbrace\left(\mbox{$\mathbf{b}^{\mathrm{R}}_{1,~n}$}^T\mathbf{\overline{x}}_m+\mbox{$\mathbf{b}^{\mathrm{R}}_{2,~n}$}^T\mathbf{\overline{x}}_{m-1} \right)^2\right\rbrace\\\label{eq_ex4_CP_app1}
&+\mathbb{E}\left\lbrace\left(\mbox{$\mathbf{b}^{\mathrm{I}}_{1,~n}$}^T\mathbf{\overline{x}}_m+\mbox{$\mathbf{b}^{\mathrm{I}}_{2,~n}$}^T\mathbf{\overline{x}}_{m-1} \right)^2\right\rbrace+\sigma_{\mathrm{n_P}}^2\\
=&\mathbb{E}\left\lbrace\mathbf{\overline{x}}_m^T\mathbf{B}_{1,~n}\mathbf{\overline{x}}_m+\mathbf{\overline{x}}_{m-1}^T\mathbf{B}_{2,~n}\mathbf{\overline{x}}_{m-1} 
+2\mathbf{\overline{x}}_m^T\mathbf{D}_{n}\mathbf{\overline{x}}_{m-1} \right\rbrace+\sigma_{\mathrm{n_P}}^2\\
=&\mathrm{Tr}\left(\mathbf{B}_{n}\mathbf{P}\right)+ 2\mathrm{Tr}\left(\mathbf{D}_{n}\mathbf{U}\right)+\sigma_{\mathrm{n_P}}^2,
\end{align}
\end{subequations}
where $\mathbf{B}_n=\mathbf{B}_{1,~n}+\mathbf{B}_{2,~n}\in\mathbb{R}^{2K\times 2K}$ with $\mathbf{B}_{1/2,~n}=\mbox{$\mathbf{b}^{\mathrm{R}}_{1/2,~n}$}\mbox{$\mathbf{b}^{\mathrm{R}}_{1/2,~n}$}^T+\mbox{$\mathbf{b}^{\mathrm{I}}_{1/2,~n}$}\mbox{$\mathbf{b}^{\mathrm{I}}_{1/2,~n}$}^T\in\mathbb{R}^{2K\times 2K}$ $\mathbf{D}_n=\mbox{$\mathbf{b}^{\mathrm{R}}_{1,~n}$}\mbox{$\mathbf{b}^{\mathrm{R}}_{2,~n}$}^T+\mbox{$\mathbf{b}^{\mathrm{I}}_{1,~n}$}\mbox{$\mathbf{b}^{\mathrm{I}}_{2,~n}$}^T\in\mathbb{R}^{2K\times 2K}$.

For the fourth-order term of CP, re-organize $y[n,~m]$ in \eqref{eq_y_[n,m]_CP} as  
\begin{subequations}
    \begin{align}
y[n,~m]=&\mbox{$\mathbf{b}^{\mathrm{R}}_{n}$}^T\mathbf{v}+\mbox{$\mathbf{b}^{\mathrm{I}}_{n}$}^T\mathbf{v}+w[n,~m],\\
    \mbox{with}~\mathbf{v}=&[{\mathbf{\overline{x}}_m}^T, ~{\mathbf{\overline{x}}_{m-1}}^T ]^T,~
    \mbox{$\mathbf{b}^{\mathrm{R/I}}_{n}$}^T=[\mbox{$\mathbf{b}^{\mathrm{R/I}}_{1,~n}$}^T, ~\mbox{$\mathbf{b}^{\mathrm{R/I}}_{2,~n}$}^T ]^T,
    \end{align}
\end{subequations}
which, following \eqref{eq_ex_1_app1}, gives 
\begin{subequations}
\begin{align}
\mathbb{E}\left\lbrace |y[n,~m]|^4\right\rbrace
=&\mathbb{E}\left\lbrace \left[\mathbf{v}^T\mathbf{B}'_{n}\mathbf{v}\right]^2\right\rbrace+6\sigma_{\mathrm{n_P}}^2\mathbb{E}\left\lbrace \mathbf{v}^T\mathbf{B}'_{n}\mathbf{v}\right\rbrace+3\sigma_{\mathrm{n_P}}^4,
\end{align}
\end{subequations}
{and}
\begin{subequations} \begin{align}
\label{eq_ex4_CP_app1}&\mathbb{E}\left\lbrace[\mathbf{v}^T\mathbf{B}'_{n}\mathbf{v}]^2\right\rbrace=\mathbb{E}\left\lbrace \mathbf{v}^T\mathbf{B}'_{n}\mathbf{v}\right\rbrace^2+\mathrm{Var}\left\lbrace\mathbf{v}^T\mathbf{B}'_{n}\mathbf{v}\right\rbrace\\
=&\mathbb{E}\left\lbrace \mathbf{v}^T\mathbf{B}'_{n}\mathbf{v}\right\rbrace^2+2\mathrm{Tr}\left(\mathbf{B}'_{n}\boldsymbol{\overline{\Sigma}}\mathbf{B}'_{n}\boldsymbol{\overline{\Sigma}}\right)+4\boldsymbol{\overline{\mu}}^T\mathbf{B}'_{n}\boldsymbol{\overline{\Sigma}}\mathbf{B}'_{n}\boldsymbol{\overline{\mu}}\\\label{eq_ex4_CP_app12}
=&\left[\mathrm{Tr}\left(\mathbf{B}_{n}\mathbf{P}\right)+2\mathrm{Tr}\left(\mathbf{D}_{n}\mathbf{U}\right)\right]^2+2\mathrm{Tr}\left(\mathbf{B}'_{n}\mathbf{\overline{P}}\mathbf{B}'_{n}\mathbf{\overline{P}}\right)-2\mathrm{Tr}\left(\mathbf{B}'_{n}\mathbf{\overline{U}}\mathbf{B}'_{n}\mathbf{\overline{U}}\right),\end{align}
\end{subequations}
with
\begin{align}
\nonumber\boldsymbol{\overline{\mu}}=&\left[\boldsymbol{\mu}^T,~\boldsymbol{\mu}^T\right]^T,~\boldsymbol{\overline{\Sigma}}=\text{diag}\left\lbrace \left[\boldsymbol{\sigma}^T,~\boldsymbol{\sigma}^T\right]^T\right\rbrace,~\mathbf{\overline{U}}=\boldsymbol{\overline{\mu}}\boldsymbol{\overline{\mu}}^T,~\mathbf{\overline{P}}=\mathbf{\overline{U}}+\boldsymbol{\overline{\Sigma}},\\
\mathbf{B}'_{n}=&\mbox{$\mathbf{b}^{\mathrm{R}}_{n}$}\mbox{$\mathbf{b}^{\mathrm{R}}_{n}$}^T+\mbox{$\mathbf{b}^{\mathrm{I}}_{n}$}\mbox{$\mathbf{b}^{\mathrm{I}}_{n}$}^T=\begin{bmatrix}
\mathbf{B}_{1,~n}&\mathbf{D}_{n}\\
\mathbf{D}^T_{n}&\mathbf{B}_{2,~n}
\end{bmatrix}.
\end{align}

If expanding the second and third term in   \eqref{eq_ex4_CP_app12} as a function of $\mathbf{U}$ and $\mathbf{P}$, we have  
\begin{align}
\nonumber&\mathrm{Tr}\left(\mathbf{B}'_{n}\mathbf{\overline{P}}\mathbf{B}'_{n}\mathbf{\overline{P}}\right)\\\nonumber
=&\mathrm{Tr}\left(\begin{bmatrix}
\mathbf{B}_{1,~n}&\mathbf{D}_n\\\mathbf{D}_{n}^T&\mathbf{B}_{2,~n}\end{bmatrix}
\begin{bmatrix}
\mathbf{P}&\mathbf{U}\\\mathbf{U}&\mathbf{P}\end{bmatrix}
\begin{bmatrix}
\mathbf{B}_{1,~n}&\mathbf{D}_n\\\mathbf{D}_{n}^T&\mathbf{B}_{2,~n}\end{bmatrix}\begin{bmatrix}
\mathbf{P}&\mathbf{U}\\\mathbf{U}&\mathbf{P}\end{bmatrix}\right)\\\nonumber
=&\mathrm{Tr}\left(\mathbf{B}_{1,~n}\mathbf{P}\mathbf{B}_{1,~n}\mathbf{P}+2\mathbf{D}_n\mathbf{P}\mathbf{D}_{n}^T\mathbf{P}+\mathbf{B}_{2,~n}\mathbf{P}\mathbf{B}_{2,~n}\mathbf{P}\right)+\\\nonumber&2\mathrm{Tr}\left(\mathbf{D}_{n}\mathbf{U}\mathbf{D}_{n}\mathbf{U}+\mathbf{B}_{1,~n}\mathbf{U}\mathbf{B}_{2,~n}\mathbf{U}\right)+2\mathrm{Tr}\left(\mathbf{B}_{1,~n}\mathbf{P}\mathbf{D}_{n}\mathbf{U}\right)\\\label{eq_BPBP1_app}
&+2\mathrm{Tr}\left(\mathbf{B}_{2,~n}\mathbf{P}\mathbf{D}^T_{n}\mathbf{U}\right)+2\mathrm{Tr}\left(\mathbf{B}_{1,~n}\mathbf{U}\mathbf{D}^T_{n}\mathbf{P}\right)+2\mathrm{Tr}\left(\mathbf{B}_{2,~n}\mathbf{U}\mathbf{D}_{n}\mathbf{P}\right),
\end{align}
and 
\begin{align}
\nonumber&\mathrm{Tr}\left(\mathbf{B}'_{n}\mathbf{\overline{U}}\mathbf{B}'_{n}\mathbf{\overline{U}}\right)\\\nonumber
=&\mathrm{Tr}\left(\begin{bmatrix}
\mathbf{B}_{1,~n}&\mathbf{D}_n\\\mathbf{D}_{n}^T&\mathbf{B}_{2,~n}\end{bmatrix}
\begin{bmatrix}
\mathbf{U}&\mathbf{U}\\\mathbf{U}&\mathbf{U}\end{bmatrix}
\begin{bmatrix}
\mathbf{B}_{1,~n}&\mathbf{D}_n\\\mathbf{D}_{n}^T&\mathbf{B}_{2,~n}\end{bmatrix}\begin{bmatrix}
\mathbf{U}&\mathbf{U}\\\mathbf{U}&\mathbf{U}\end{bmatrix}\right)\\
=&\mathrm{Tr}\left( \mathbf{E}_{n}\mathbf{U}\mathbf{E}_{n}^T\mathbf{U}\right),
\end{align}
where $\mathbf{E}_n=\mathbf{E}_{1,~n}+\mathbf{E}_{2,~n}^T=\left(\mbox{$\mathbf{b}^{\mathrm{R}}_{1,~n}$}+\mbox{$\mathbf{b}^{\mathrm{R}}_{2,~n}$}\right)\left(\mbox{$\mathbf{b}^{\mathrm{R}}_{1,~n}$}+\mbox{$\mathbf{b}^{\mathrm{R}}_{2,~n}$}\right)^T+\left(\mbox{$\mathbf{b}^{\mathrm{I}}_{1,~n}$}+\mbox{$\mathbf{b}^{\mathrm{I}}_{2,~n}$}\right)\left(\mbox{$\mathbf{b}^{\mathrm{I}}_{1,~n}$}+\mbox{$\mathbf{b}^{\mathrm{I}}_{2,~n}$}\right)^T$ is positive semi-definite with  $\mathbf{E}_{1,~n}=\mathbf{B}_{1,~n}+\mathbf{D}_n$ and $\mathbf{E}_{2,~n}=\mathbf{B}_{2,~n}+\mathbf{D}_n$.

\bibliographystyle{IEEEtran}
\bibliography{IEEEabrv, references}
\end{document}

%% file: notation_new.tex
\def\nb0{{\mathbf{0}}}
\def\nb1{{\mathbf{1}}}

\def\ncalC{{\mathcal{C}}}

\def\ncalN{{\mathcal{N}}}

\newtheorem{ndef}{Definition}
\newtheorem{nrem}{Remark}

\newtheorem{eg}{Example}

